\documentclass[aps,pra,reprint,superscriptaddress,showkeys,amsmath,amssymb,longbibliography]{revtex4-1}
\usepackage[english]{babel}
\usepackage{amsmath,amssymb,bbm,mathrsfs,bm,braket,color,graphicx,comment,amsfonts,dsfont}
\usepackage[colorlinks, citecolor=blue, urlcolor=blue]{hyperref}
\usepackage[normalem]{ulem}
\usepackage{multirow}
\usepackage[normalem]{ulem}

\begin{document}
\title{Supermetal-insulator transition in a non-Hermitian network model}

\author{Hui Liu}
\affiliation{IFW Dresden and W{\"u}rzburg-Dresden Cluster of Excellence ct.qmat, Helmholtzstrasse 20, 01069 Dresden, Germany}

\author{Jhih-Shih You}
\affiliation{Department of Physics, National Taiwan Normal University, Taipei 11677, Taiwan}

\author{Shinsei Ryu}
\affiliation{Department of Physics, Princeton University, Princeton, New Jersey, 08540, USA}

\author{Ion Cosma Fulga}
\affiliation{IFW Dresden and W{\"u}rzburg-Dresden Cluster of Excellence ct.qmat, Helmholtzstrasse 20, 01069 Dresden, Germany}

\begin{abstract}
We study a non-Hermitian and non-unitary version of the two-dimensional Chalker-Coddington network model with balanced gain and loss. 
This model belongs to the class D$^\dagger$ with particle-hole symmetry$^\dagger$ and hosts both the non-Hermitian skin effect as well as exceptional points.
By calculating its two-terminal transmission, we find a novel contact effect induced by the skin effect, which results in a non-quantized transmission for chiral edge states.
In addition, the model exhibits an insulator to `supermetal' transition, across which the transmission changes from exponentially decaying with system size to exponentially growing with system size.
In the clean system, the critical point separating insulator from supermetal is characterized by a non-Hermitian Dirac point that produces a quantized critical transmission of $4$, instead of the value of $1$ expected in Hermitian systems.
This change in critical transmission is a consequence of the balanced gain and loss.
When adding disorder to the system, we find a critical exponent for the divergence of the localization length $\nu\approx 1$, which is the same as that characterizing the universality class of two-dimensional Hermitian systems in class D. 
Our work provides a novel way of exploring the localization behavior of non-Hermitian systems, by using network models, which in the past proved versatile tools to describe Hermitian physics.
\end{abstract}
\maketitle

\section{Introduction}
\label{sec:introduction}

Topological insulators are phases of matter in which an insulating bulk coexists
with robust, conducting edge states~\cite{Hasan2010rmp, 2016RvMP...88c5005C}.
The conducting edges are protected by a topological invariant defined from the bulk band topology, a manifestation of the so-called bulk-edge correspondence.
In Hermitian systems, the edge states of two-dimensional topological insulators are characterized by a quantized conductance.
One can expect that, if Hermiticity is broken~\cite{Bergholtz2021rmp, Ashida2020, MartinezAlvarez2018}, the topological structure could be significantly changed due to the complex eigenvalues, and the imaginary part of eigenvalues could cause an amplification or a decay of the quantized conductance.

Non-Hermiticity arises naturally as an effective description of a wide range of systems, including electronic systems with a finite lifetime of quasiparticles~\cite{Kozii2017, Shen2018, Papaj2019, Nagai2020}, as well as photonic systems under the influence of radiative loss~\cite{Feng2017, ElGanainy2018, Ozawa2019,zdemir2019}.  
One of the more prominent phenomena present in non-Hermitian systems is the emergence of exceptional points~(EPs)~\cite{Bender1999, Heiss2012, Miri2019, Bergholtz2021rmp}, where eigenvalues and eigenstates of the non-Hermitian matrix coalesce. 
Analogous to the surface Fermi arcs of Weyl semimetals, each pair of exceptional points is connected by an open-ended bulk Fermi arc in the real (or imaginary) energy spectrum~\cite{Kozii2017, Malzard2018, Zhou2018, Carlstreom2018}. 
Another unique phenomenon present in non-Hermitian topology is the breakdown of the bulk-edge correspondence~\cite{Yao2018}. 
In certain types of non-Hermitian systems, such as when the hoppings are non-reciprocal~\cite{Hatano1996, Hatano1997}, the presence of open boundaries can cause all eigenmodes to become localized on the edge and their eigenvalues to be completely different from those of the corresponding system with periodic boundary conditions.
This phenomenon is called the non-Hermitian skin effect~\cite{Lee2016, Kunst2018, Xiong2018, Yao2018, Yao2018a, Yokomizo2019, Zhang2020, Longhi2020, Li2020, zhang2021}, and has by now been experimentally observed in a variety of systems~\cite{Xiao2020, Helbig2020, Ghatak2020, Weidemann2020}. 

Recently, several works have examined the conductance of two-dimensional non-Hermitian topological insulators~\cite{Chenyu2018, Philip2018, Wang2019, Hirsbrunner2019, Groenendijk2020}. 
However, the correspondence between this conductance and the non-Hermitian band topology, the exceptional points, and the non-Hermitian skin effect remains largely unexplored. 
A further question is how the `topological semimetal' phase characterized by the bulk Fermi arc evolves during the localization-delocalization transitions produced by disorder.

In our work, we address these questions by turning to one of the well-known tools used to characterize the localization behavior of Hermitian systems: the network model.
By introducing balanced gain and loss to the original Chalker-Coddington~(CC) network model~\cite{Chalker1988}, we show that the resulting non-unitary network can realize both EPs and the non-Hermitian skin effect.
We compute the two-terminal transmission and find that there is a contact effect induced by non-Hermiticity, in the sense that the transmission is influenced by the lead geometry and orientation.
While reminiscent of the contact resistance characterizing transport measurements in conventional, Hermitian mesoscopic samples~\cite{datta1995, imry1997}, the non-Hermitian contact effect has drastic consequences on the system's transport properties.
Due to this contact effect, the transmission through the chiral edge states is influenced by the non-Hermitian skin effect and becomes non-quantized, even when gain and loss are balanced.

Furthermore, we find that the two-terminal transmission shows a transition between gapped and gapless phases.
At the phase boundary, the non-Hermitian Dirac point possesses a quantized transmission probability equal to $4$, instead of the usual value of $1$, characteristic of Hermitian Dirac cones.
By applying a finite-size scaling fit to the transmission, we find such a bulk `topological semimetal' behaves like a `supermetal' in which the transmission increases exponentially with system size.
The fit results show a critical exponent $\nu\approx 1$ for the divergence of the localization length, which means the critical exponent $\nu$ for Hermitian class D systems~\cite{Altland1997,Medvedyeva2010,Ludwig1994} is also valid for the non-Hermitian class D$^\dagger$~\cite{Kawabata2019b}.
However, as far as we are able to deduce from our numerical results, the critical transmission seems to no longer show a universal value in this non-unitary system.

The rest of this work is organized as follows. In Sec.~\ref{sec:model} we introduce a modified version of the Chalker-Coddington model, by adding balanced gain and loss to the system.
We briefly recapitulate the properties of the original model (Sec.~\ref{sec:top_phase}), and then describe the new features introduced by non-Hermiticity: exceptional points (Sec.~\ref{sec:ep}) and the non-Hermitian skin effect (Sec.~\ref{sec:se}).
In Sec.~\ref{sec:ce} we describe the procedure used to determine the transport properties of the network model, highlighting the emergence of the non-Hermitian contact effect in Sec.~\ref{subsec:contact}, as well as discussing parameter regions in which infinite amplification loops cause numerical instabilities (Sec.~\ref{sec:ni}).
The transport properties of the clean system are analyzed in Sec.~\ref{sec:transport_clean}, where we show that the non-Hermitian Dirac cone is characterized by a quantized conductance different from its Hermitian counterpart.
Turning to the disordered system in Sec.~\ref{sec:disorder}, we show its phase diagram, determine its critical exponent to be the same as that of Hermitian systems in class D, and provide a heuristic argument to justify this finding.
Finally, we conclude and discuss directions for future research in Sec.~\ref{sec:conclusion}.

\section{Network model}
\label{sec:model}

\begin{figure}
\centering
\includegraphics[width=\columnwidth]{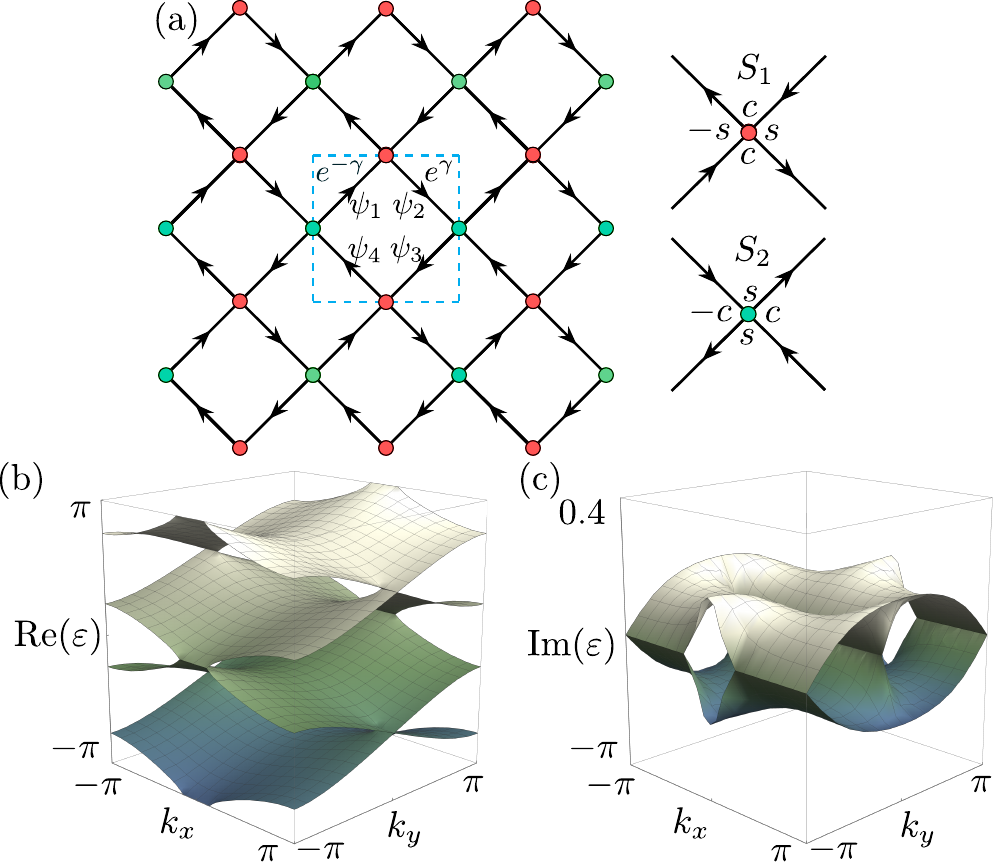}
\caption{(a) Illustration of the non-unitary network model. 
$\psi_i$ ($i = 1$, $2$, $3$, $4$) represents a propagating mode inside the unit cell (blue dashed square). 
Red and green nodes correspond to two types of $2\times 2$ scattering matrices $S_1$ and $S_2$ with $s\equiv \sin\alpha$ and $c\equiv \cos\alpha$, respectively. 
(b) and (c) are the real and imaginary eigenphase spectra in momentum space, respectively. 
The parameters are $\alpha=\pi/4$ and $\gamma=0.5$.
\label{fig: set_up}}
\end{figure}

The Chalker-Coddington model is a network of unidirectional modes, similar to the chiral edge modes of the quantum Hall effect \cite{Chalker1988}.
These modes comprise the links of the network and scatter into each other at the nodes of the network, which are assumed to form a periodic pattern in real space.
Each one of the scattering events is modeled using a scattering matrix that connects the probability amplitudes of incoming and outgoing modes.
Therefore, the CC network model is described by an array of scattering matrices that encodes the full propagation of states through the system.

We start from the CC network model with two different types of scattering nodes $S_1$ and $S_2$ inside one unit cell~[see Fig.~\ref{fig: set_up}(a)]. 
Each node transfers two incoming modes to two outgoing modes, the degree of mixing between them being parameterized by the mixing angle $\alpha \in (-\pi, \pi]$.
Thus, as shown Fig.~\ref{fig: set_up}(a), a mode entering from the top into a node of type $S_1$ has a probability amplitude $\cos\alpha$ of turning clockwise, and a probability amplitude $\sin\alpha$ of turning counter-clockwise.
The relative minus signs describing scattering events on the nodes $S_1$ and $S_2$ are introduced in order to ensure current conservation, that is to ensure that the scattering matrices describing the two types of node are unitary.

To model a non-Hermitian and non-unitary system, we introduce gain and loss into the CC model.
After each scattering event, one of the two outgoing modes is either amplified or attenuated, by $e^{\gamma}$ and $e^{-\gamma}$, respectively, as shown in Fig.~\ref{fig: set_up}(a).
The resulting network model is the same as that introduced in Ref.~\cite{Hwsc_model}, which used it primarily to study quantum pumping.

The behavior of the network can be studied by defining a network wavefunction, $\Phi$, describing the amplitude of a state in each of the unidirectional links of the system.
The process by which different links scatter into each other is encoded in the scattering amplitudes of the different nodes, which together form a so-called Ho-Chalker operator \cite{Ho1996} that acts on the network wavefunction:
\begin{equation}
\Phi(t+1) = {\cal S} \Phi(t).
\end{equation}
Thus, the network can be understood as describing a discrete time-evolution process, parameterized by the integer time $t$, similar to periodically-driven, or Floquet systems.
Here, however, $t$ labels the state of the network wavefunction in between different scattering events.

By analogy with Floquet systems, an eigenstate of the network model can be thought of as a stationary state, one which retains its shape after multiple scattering events. 
Using the translation symmetry of the model in Fig.~\ref{fig: set_up}(a), it is advantageous to go to momentum space, $\textbf{k}=(k_x,k_y)$, where the wavefunction $\Phi(\textbf{k})=(\psi_1(\textbf{k}),\psi_3(\textbf{k}),\psi_2(\textbf{k}),\psi_4(\textbf{k}))^{T}$ has four components, since there are only four propagating states inside a unit cell.
The eigenstates of the network can then be found from the Ho-Chalker operator in momentum space~\cite{Ho1996}
\begin{eqnarray}
\mathcal{S}(\textbf{k})\Phi(\textbf{k})=e^{i\varepsilon(\textbf{k})}\Phi(\textbf{k}),\label{eq:hc_momentum}
\end{eqnarray}
with $\varepsilon(\textbf{k})$ describing the eigenphases of the network, similar to the quasienergies characterizing Floquet systems.
The resulting bandstructure is shown in Figs.~\ref{fig: set_up}(b, c).

So far, this description is identical to the one commonly used to study Hermitian systems using network models.
The main difference introduced in this work is that, due to the gain and loss added to the network model, the Ho-Chalker operator is no longer unitary.
Instead, $\mathcal{S}(\textbf{k})=\mathcal{U}_{\rm nu}\mathcal{S}_{\rm u}(\textbf{k})$, with $\mathcal{U}_{\rm nu}=\text{diag}[e^{-\gamma},1,e^{\gamma},1]$ and $\mathcal{S}_{\rm u}(\textbf{k})$ is the unitary Ho-Chalker operator for the original CC network model
\begin{eqnarray}
\hspace{-2mm}\mathcal{S}_{\rm u}(\textbf{k})=\begin{pmatrix}0&0&\sin\alpha e^{-ik_x}&\cos\alpha\\0&0&-\cos\alpha&\sin\alpha e^{ik_x}\\\cos\alpha&\sin\alpha e^{ik_y}&0&0\\-\sin\alpha e^{-i k_y}&\cos\alpha&0&0\end{pmatrix}.\nonumber\\
\label{eq: ho_chalker_operator}
\end{eqnarray}

In the language of discrete time evolution, a unitary Ho-Chalker operator corresponds to the dynamics generated by a Hermitian Hamiltonian.
Here, the Ho-Chalker operator is not unitary if $\gamma\neq 0$, and thus the dynamics it describes corresponds to a time evolution governed by a non-Hermitian Hamiltonian.
As a result, eigenphases are now complex, as shown in Figs.~\ref{fig: set_up}(b, c), in contrast to the real quasienergies expected for a periodically-driven Hermitian model.
This is the sense in which the modified CC model we study describes the behavior of non-Hermitian systems.

In the unitary limit~($\gamma=0$), the system belongs to class D in the Altland Zirnbauer (AZ) classification \cite{Altland1997}; it is the well-known Cho-Fisher model~\cite{Cho_Fisher}.
It obeys particle-hole symmetry~(PHS) because of $\mathcal{S}(\textbf{k})=\mathcal{S}^{*}(-\textbf{k})$.
This means that for any eigenstate at an eigenphase $\varepsilon$ and momentum $\textbf{k}$, there must exist an eigenstate at $-\varepsilon$ and $-\textbf{k}$.
Furthermore, we note that there exists another symmetry, the phase-rotation symmetry $\mathcal{U}_p\mathcal{S}(\textbf{k})\mathcal{U}_p^{-1}=e^{-i\pi}\mathcal{S}(\textbf{k})$ with $\mathcal{U}_p=-\sigma_z\otimes \sigma_0$~\cite{Delplace2017} ($\sigma$ are the conventional Pauli matrices).
Because of phase-rotation symmetry, for any state at eigenphase $\varepsilon$ there must exist another state at $\varepsilon+\pi$, meaning that the spectrum of the system repeats twice in the full interval $\varepsilon\in[-\pi,\pi)$.

When switching on gain and loss, the unitarity (and therefore current conservation) is broken: $\mathcal{S}(\textbf{k})\mathcal{S}^{\dagger}(\textbf{k})\neq 1$. 
However, the system still inherits $\mathcal{S}(\textbf{k})=\mathcal{S}^{*}(-\textbf{k})$ and the phase-rotation symmetry, since $\mathcal{U}_{\rm nu}$ is a real diagonal matrix and obeys $[\mathcal{U}_{\rm nu},\mathcal{U}_{p}]=0$.
According to the non-Hermitian $38$-fold symmetry classification~\cite{Kawabata2019b}, such a system belongs to the class D$^\dagger$ with PHS$^\dagger$~(see Appendix~\ref{app:sym_eo}). 
In this sense, for any eigenstate at $\varepsilon$ and $\textbf{k}$, there is an eigenstate at $-\varepsilon^{*}$ and $-\textbf{k}$.

\section{Hermitian limit}
\label{sec:top_phase}

\begin{figure}
\centering
\includegraphics[width=1\columnwidth]{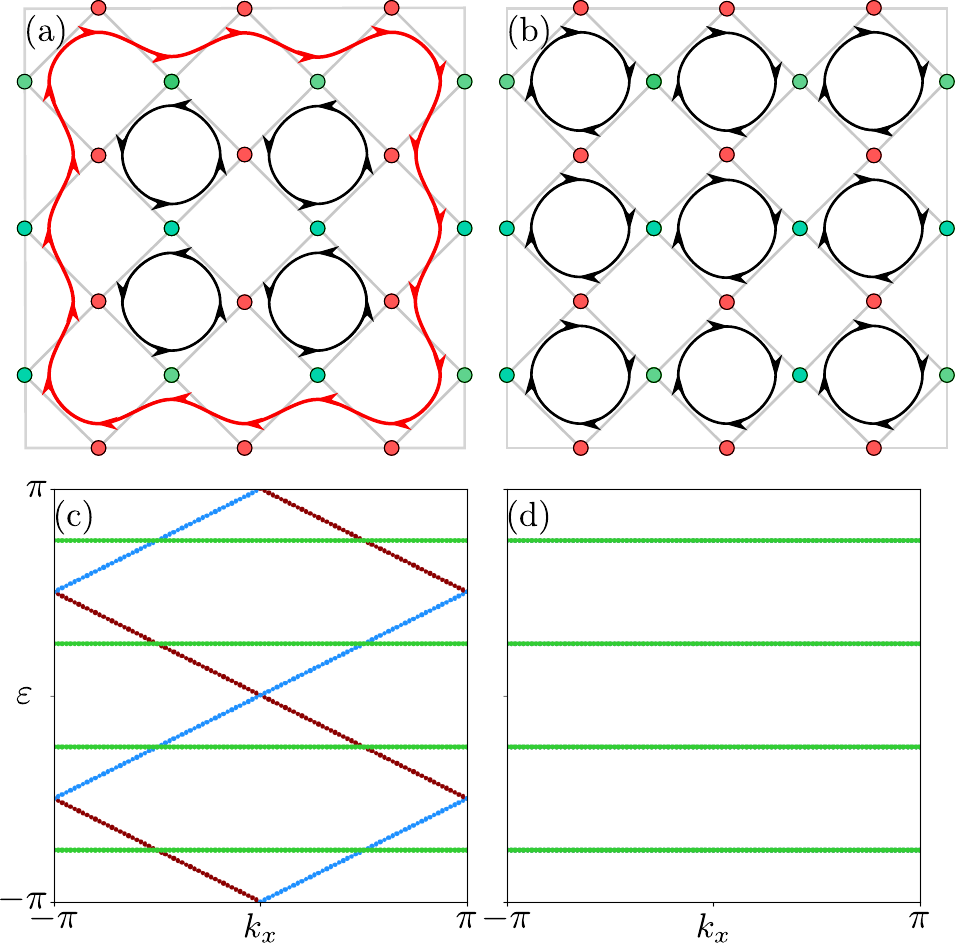}
\caption{Topological phase limit with $\alpha=\pi/2$ (a) and trivial phase limit with $\alpha=0$ (b). 
(c-d) are their corresponding ribbon geometry eigenphase spectra. 
The ribbon consists of 3 unit cells in the vertical direction and is infinite in the horizontal direction. 
The green, blue, and red color indicate eigenstates localized at the bulk, the top edge, and the bottom edge, respectively.
\label{fig: decoupled_limit}}
\end{figure}

As a model to simulate the integer quantum Hall transition, the Hermitian ($\gamma=0$) CC network supports a topological distinction between the trivial phase ($\alpha < |\pi/4|$) and the topological phase [$\alpha\in (\pi/4, 3\pi/4)$ or $\alpha\in (-3\pi/4, -\pi/4)$]. 
Such phases can be seen intuitively by the decoupled limits, $\alpha=0,\pi/2$, in which all propagating modes turn either clockwise or counter-clockwise with unit probability at the scattering nodes [see Fig.~\ref{fig: decoupled_limit}(a, b)]. 
Here, we obtain the finite network by setting a hard wall boundary where the propagating modes are fully reflected.
As can be seen, all of the unidirectional modes on the links of the network form closed loops which are decoupled from each other, meaning that all bulk states are localized. 
However, the different pattern of closed loops for $\alpha=0$ and $\alpha=\pi/2$ means that in the latter case there is a chiral edge mode encircling the system perimeter.
This is a topologically protected edge state, similar to that present in the integer quantum Hall effect.

The localized bulk states and the chiral edge mode can also be seen in the eigenphase spectrum of the CC network obtained in a ribbon geometry, as shown Fig.~\ref{fig: decoupled_limit}(c, d). 
In both cases, the bulk states are dispersionless and located at $\varepsilon = \pm\pi/4$ and $\varepsilon=\pm3\pi/4$. 
In the nontrivial phase of  Fig.~\ref{fig: decoupled_limit}(c), however, there appear chiral boundary modes on the ribbon edges, which wind both in momentum and eigenphase. 
Since the Chern number of a bulk band is equal to the net difference between the number of chiral edge states above and below the band, we can see that all bulk bands are trivial.
As such, the topological phase of the CC model is analogous to so-called anomalous Floquet topological phases \cite{Rudner_PRX, Titum_PRX, Maczewsky2017, Mukherjee2017}.

The phase transitions between the topological and trivial phases occur by means of gap closings and reopenings between eigenphase bands.
In the Hermitian CC network, such gapless points occur at $\alpha=\pm\pi/4$, where propagating states have an equal probability of turning clockwise and counter-clockwise at each node \cite{Chalker1988, KRAMER2005211}.
The result is the formation of Dirac points between the bulk bands, which appear at $\varepsilon=0$, $\pm\pi/2$, and $\pi$.
We note that there exist other gapless points at $\alpha=\pm3\pi/4$, which are also characterized by a Dirac eigenphase spectrum, and which mark a transition to a weak topological phase if further increasing $|\alpha|$. 

\section{Exceptional points}
\label{sec:ep}

When turning on non-Hermiticity, $\gamma\neq0$, the Dirac points of the network model split into pairs of EPs at which complex eigenphase bands coalesce.
To find these points, we use the $\pi$-phase rotation symmetry of the network together with the condition imposed by balanced gain and loss, $\text{det}[\mathcal{S}(\textbf{k})]=1$.
The precondition for EPs becomes 
\begin{eqnarray}
e^{2i\varepsilon}=\pm 1.
\end{eqnarray}
This equation shows that there could exist EPs at $\varepsilon=0$, $\pm\pi/2$, and $\pi$, as shown in Fig.~\ref{fig: set_up}(b-c). 
We focus on EPs at $\varepsilon=0$. 
By solving $\text{Det}[\mathcal{S}(\textbf{k})-I]=0$, we analytically identify EPs $\textbf{k}_{\text{EP}}$  at
\begin{eqnarray}\label{eq:eps}
\hspace{-5mm}k_x&=&\pm\arccos(\frac{2}{\sin 2\alpha}+\cosh\gamma), k_y=0, \alpha\in (\pi/2,\pi),\\
\hspace{-5mm}k_x&=&\pm\arccos(\frac{2}{\sin 2\alpha}-\cosh\gamma), k_y=\pi, \alpha\in (0,\pi/2). 
\end{eqnarray}
Since $\cos k_x \in [-1, 1]$ for real $k_x$  we find that
\begin{eqnarray}
\frac{2}{\sin 2\alpha}+\cosh\gamma&=&\pm 1,\quad \alpha\in (\pi/2,\pi),\label{eq:boundary_1}\\
\frac{2}{\sin 2\alpha}-\cosh\gamma&=&\pm 1,\quad \alpha\in (0,\pi/2), \label{eq:boundary_2}
\end{eqnarray}
determines the boundary between gapped and gapless systems with a pair of EPs.
This indicates that, as $\gamma$ increases away from zero, the $\varepsilon=0$ gap first closes to form a Dirac point, after which this Dirac point splits into a pair of EPs.
The latter move in momentum space along the $k_x$-direction as $\gamma$ is further increased, eventually annihilating at $k_x=\pi$.

During their evolution, the EPs are connected by a line for which $\text{Re}(\varepsilon)=0$ [see Fig.~\ref{fig: set_up}(b)], known as a bulk Fermi arc, which is analogous of the Fermi arcs of three-dimensional Hermitian Weyl semimetals~\cite{rmp_weyl, Wan_weyl, Binghai_weyl, Burkov_weyl}.
This bulk Fermi arc is a topological consequence of the EPs.
When forming a closed loop $s^1$ in momentum space that encircles one of the EPs, the presence of a bulk Fermi arc is connected to a nonzero winding number~\cite{Kawabata2019}
\begin{eqnarray}
\mathcal{W}=\oint_{s^1}\frac{d\textbf{k}}{2\pi i}\cdot \nabla_{\textbf{k}}\text{log}[\text{det}[\mathcal{S}(\textbf{k})-e^{i\varepsilon(\textbf{k}_{\text{EP}})}]].\label{eq: winding}
\end{eqnarray}
We find that the two EPs have $\mathcal{W}=\pm 1$, and are thus topologically protected~(see Appendix \ref{app:winding}).

The formation of EPs at the transition between trivial and topological phases can also be deduced from the long-wavelength limit of the network model.
By expanding the Ho-Chalker operator $\mathcal{S}(\textbf{k})$ around $\varepsilon=0$, $\alpha=\pi/4$, and $\gamma=0$, we obtain an effective Hamiltonian describing a non-Hermitian Chern insulator, as expected.
The later takes the form 
\begin{eqnarray}
  \label{dirac non hermitian Chern ins}
\mathcal{H}=p_x\frac{\sigma_{-}}{2}+(p_y-i\gamma)\frac{\sigma_{+}}{2}+m\sigma_y,
\end{eqnarray}
with $\sigma_{\pm}=\sigma_z\pm \sigma_x$ and $m$ being the mass term (see Appendix \ref{app:longwavelength_limit} for more details).

\section{Skin effect}
\label{sec:se}

\begin{figure}
\centering
\includegraphics[width=1\columnwidth]{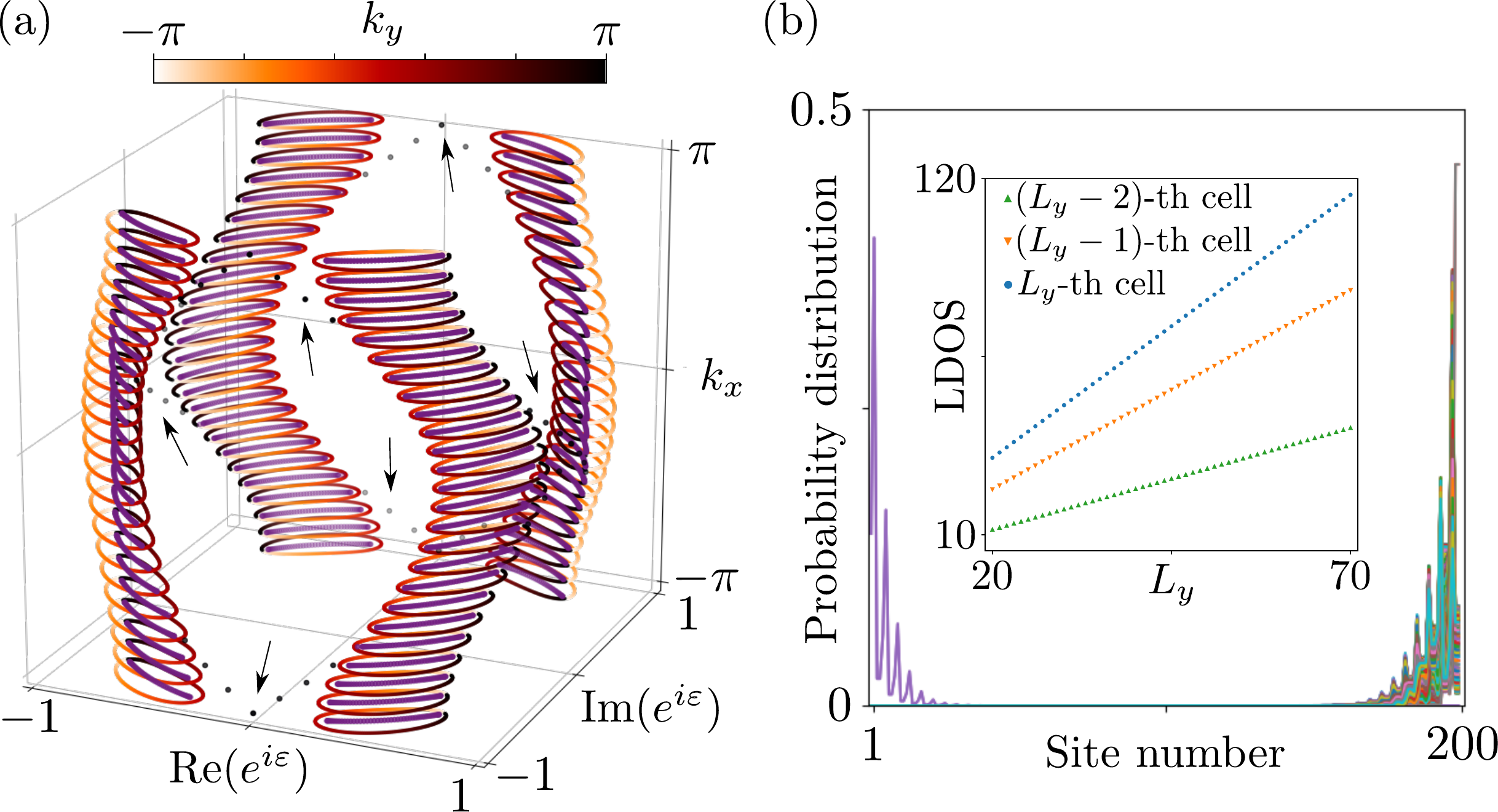}
\caption{(a) The eigenphases obtained in a ribbon geometry with a width of of 50 unit cells (purple color) are circled by the eigenphases of the infinite system. 
Outside of these winding contours, there appear chiral edge modes located on the ribbon boundaries (black points, shown by arrows). 
(b) The probability distribution of all the eigenvectors with open boundary conditions. 
All bulk states are pushed to the bottom boundary, but the top edge mode remains unaffected.
The inset shows the sum of probability densities for all states on the bottom three unit cells, which we label `LDOS' by analogy to Hermitian systems. 
The summed probability density increases linearly with the width of the ribbon, indicating that an extensive number of states are present on the bottom boundary.
The plots are with $\alpha=0.35\pi$, $\gamma=0.5$, and for panel (b) we chose $k_x=0.05$. 
\label{fig: skin_effect}}
\end{figure}

In addition to exceptional points, the network model also shows a non-Hermitian skin effect. 
By solving $\text{det}[\mathcal{S}(\textbf{k})-e^{ i \varepsilon ( \textbf{k} )}]=0$, we obtain
\begin{eqnarray}
e^{i\varepsilon(k_x, k_y)}&=&e^{-i\varepsilon^{*}(k_x,-k_y)},\label{eq: fix_kx}\\
e^{i\varepsilon(k_x, k_y)}&=&e^{i\varepsilon(-k_x,k_y)}. \label{eq: fix_ky}
\end{eqnarray} 
When $\gamma\neq0$, such that eigenphases are complex, the above equations suggest that for each fixed $k_x$ the eigenphase bands $\varepsilon$ wind in the complex plane as a function of $k_y$.
When an open boundary condition~(OBC) is imposed along the $y$-direction, thus forming a ribbon, we find that eigenphases form arcs located inside the winding contours of the infinite system spectrum, as shown in Fig.~\ref{fig: skin_effect}(a).
For an infinite system, the eigenphase bands at fixed $k_x$ have a winding number $\mathcal{W}=1$ as a function of $k_y$.
As a consequence \cite{Okuma2020}, the eigenvectors with OBC are localized at the bottom boundary, giving rise to the non-Hermitian skin effect~[see Fig.~\ref{fig: skin_effect}(b)].

Interestingly, while all bulk states become localized on the bottom boundary, the chiral edge modes remain unaffected.
This can be seen when looking at the mode localized on the top boundary [see Fig.~\ref{fig: skin_effect}(b)], which retains its position despite the non-Hermitian skin effect.
The immunity of this mode is due to the fact that it is present at eigenphases outside of the winding contours of the bands of the infinite system, for which the winding number vanishes.
This difference in edge versus bulk mode behavior is similar to the one found in Ref.~\cite{Kawabata2018}, which examined a non-Hermitian Hamiltonian realizing the quantum Hall effect.

We find that it is possible to control if the top edge mode is influenced by the non-Hermitian skin effect or not, depending on the existence of the bulk exceptional points.
If $\gamma$ is increased until the EPs appear at $k_x=0$, the connectivity of the winding eigenphase bands changes, and the the edge modes now reside in a region of winding number $\mathcal{W}=1$.
In this case, the top boundary mode is pushed to the bottom edge together with all other states (see Appendix \ref{app:se_es}). 

Note that the non-Hermitian skin effect is only present in the $y$-direction. 
As can be seen in  Fig.~\ref{fig: skin_effect}(a), the eigenphases of the infinite system do not wind as a function of $k_x$ when $k_y$ is fixed.
Therefore, if considering a ribbon which is finite in the $x$-direction and infinite along $y$, no accumulation of states would occur on the system boundaries.
In this case, the probability distribution of states (not shown) resembles that of a conventional, Hermitian Chern insulator. 
There is one chiral edge mode on each boundary of the system, and the other states are spread uniformly throughout the system bulk.
The presence of the skin effect only along one direction can be understood from the structure of the gain and loss terms in real space, as shown in Fig.~\ref{fig: set_up}(a).
For $\gamma>0$, downward moving modes are amplified, $e^\gamma$, and upward moving modes are attenuated, $e^{-\gamma}$, consistent with the accumulation of bulk states on the bottom boundary, shown in Fig.~\ref{fig: skin_effect}(b).
As such, it is possible to have bulk states accumulate on the top boundary by changing the sign of $\gamma$.
In contrast, left moving or right moving modes experience equal amounts of gain and loss on average, regardless of the sign of $\gamma$.

It is however possible to modify the system such that it shows a skin effect in both the $x$ and $y$ directions.
This leads to a so-called a second-order skin effect~\cite{Kawabata_prb2020}, in which all of the bulk states are pushed to a corner of the system. 
However, one needs to slightly change the unit cell, see Appendix~\ref{app:hose}.

\section{Two-terminal geometry}
\label{sec:ce}

To build the connection between the non-Hermitian band topology and transport, we study the two-terminal transmission by attaching two leads to the boundary of the network model along the horizontal~(h) or the vertical~(v) direction.
The leads are formed from incoming and outgoing chiral modes that have the same structure as those present on the links of the network model.
They are attached to the boundary nodes of the system in such a way as to preserve the shape of the node scattering matrices $S_{1,2}$ at those boundaries.

In general, the two-terminal scattering matrix takes the form
\begin{eqnarray}
S_{\text{h,v}}(\gamma)=\begin{pmatrix}
\mathfrak{r}_{\text{h,v}}(\gamma)&\mathfrak{t}'_{\text{h,v}}(\gamma)\\\mathfrak{t}_{\text{h,v}}(\gamma)&\mathfrak{r}’_{\text{h,v}}(\gamma)
\end{pmatrix},
\end{eqnarray}
where $\mathfrak{r}^{(\prime)}$ and $\mathfrak{t}^{(\prime)}$ are blocks containing the probability amplitudes for modes that are reflected back into the same lead, or transmitted between the leads, respectively.

If the system is Hermitian, the transmission is given by $G_{\text{h,v}}(0)=\text{tr}(\mathfrak{t}'_{\text{h,v}}(0)\mathfrak{t}'^{\dagger}_{\text{h,v}}(0))=\text{tr}(\mathfrak{t}_{\text{h,v}}(0)\mathfrak{t}^{\dagger}_{\text{h,v}}(0))$, where tr denotes the trace.
When Hermiticity is broken, $\text{tr}(\mathfrak{t}'_{\text{h,v}}(\gamma)\mathfrak{t}'^{\dagger}_{\text{h,v}}(\gamma))$ could be different from $\text{tr}(\mathfrak{t}_{\text{h,v}}(\gamma)\mathfrak{t}^{\dagger}_{\text{h,v}}(\gamma))$.
See Appendix~\ref{app:sym_eo} for a derivation of the constraints imposed on the scattering matrix in different non-Hermitian symmetry classes.

\subsection{Contact effect}
\label{subsec:contact}

\begin{figure}
\centering
\includegraphics[width=1\columnwidth]{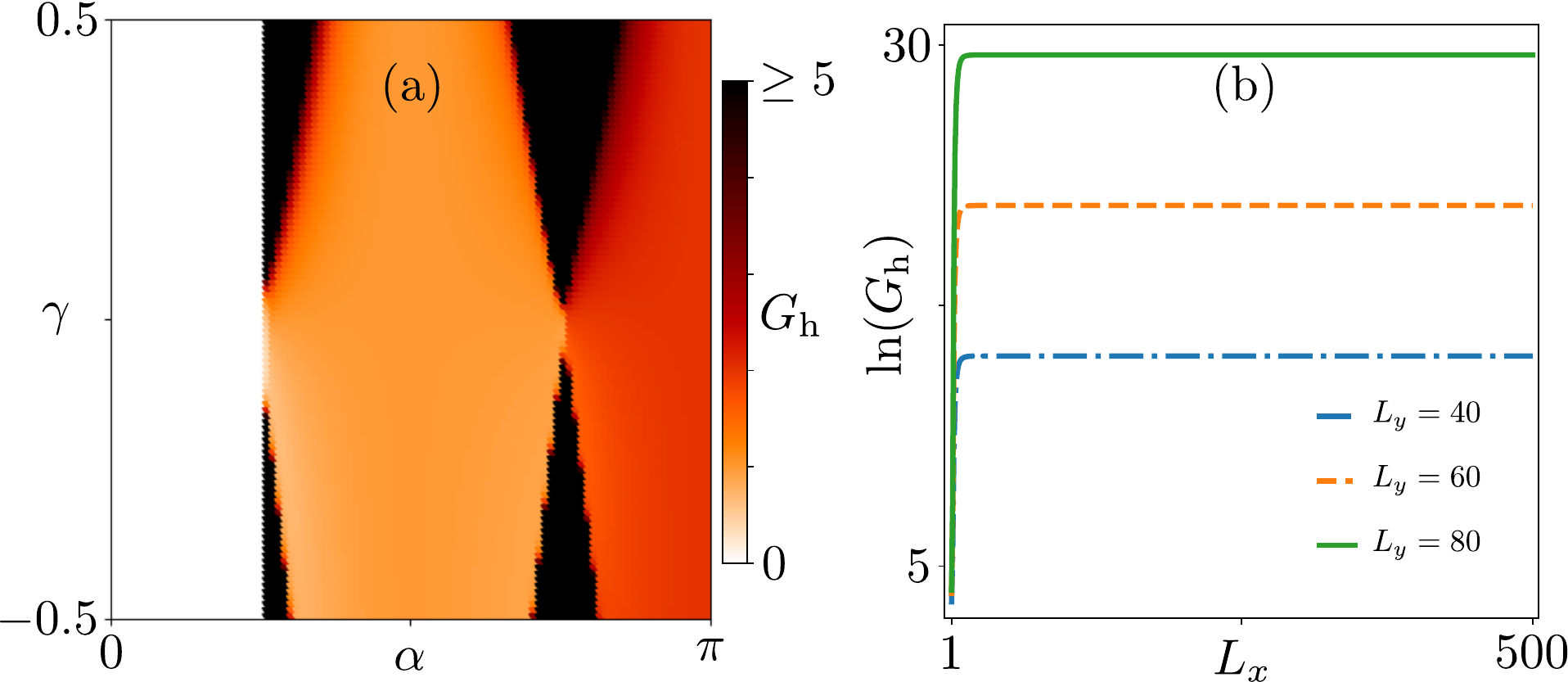}
\caption{(a) The horizontal transmission $G_{\text{h}}$ with OBC as a function of $\alpha$ and $\gamma$. 
The system size is $L_x=3000$ and $L_y=60$. (b) $G_{\text{h}}$ for different $L_y$ with open boundary condition as a function of $L_x$. 
This plot is with $\alpha = 0.3\pi$ and $\gamma=0.5$.
\label{fig: horizontal_transmission}}
\end{figure}

First, we focus on the horizontal transmission $G_{\text{h}}$ with OBC along the $y$-direction, for which the system shows a non-Hermitian skin effect.
Since the eigenphase spectrum is real with OBC~[all OBC eigenvalues are located on the unit circle in the complex plane in Fig.~\ref{fig: skin_effect}(a)], we would expect the transmission to be consistent with that found in the Hermitian limit, in which $G_{\text{h}}=0$ for the trivial phase and $G_{\text{h}}=1$ for the strong topological phase with chiral edge states.
Surprisingly, we find the non-Hermitian network shows a very different behavior.
While the transmission is zero for the trivial phase, it is not quantized when chiral edge states are present~[Fig. \ref{fig: horizontal_transmission}(a)].
$G_{\text{h}}$ can reach beyond $5$ even for small $|\gamma|$ when $\alpha$ approaches the phase transition points~($\pi/4$ or $3\pi/4$) and returns back to around $1$ if $\alpha$ is far away from the phase transition points.

These phenomena point to the presence of a contact effect when attaching leads.
The width of the chiral edge states along the $y$-direction is not exactly $0$ [as shown in Fig.~\ref{fig: skin_effect}(b)], except in the decoupled limit with $\alpha=\pi/2$, so amplification or loss can occur inside the wave packet of the chiral edge states.
The reason for this amplification and attenuation is the structure of the network model itself, in which downward moving modes acquire a factor $e^\gamma$ and all upward moving modes acquire a factor $e^{-\gamma}$, as shown in Fig.~\ref{fig: set_up}(a).

To quantify the presence of the contact effect in the language of scattering matrices, we label incoming and outgoing modes of both the left and right lead by their real-space position along the $y$-direction.
Then, the transmission probability from the incoming mode $i$ in the left lead to the outgoing mode $j$ in the right lead lead is multiplied by $e^{(i-j)\gamma}$, which can be understood as a result of the skin effect with suppression and magnification in opposite directions.
In fact, it can be verified that the total transmission matrix $\mathfrak{t}_{\text{h}}(\gamma)$ can be related to the transmission matrix of the Hermitian model $\mathfrak{t}_{\text{h}}(0)$ as
\begin{eqnarray}
\mathfrak{t}_{\text{h}}(\gamma)=\mathcal{U}_a\mathfrak{t}_{\text{h}}(0)\mathcal{U}_l,\label{eq:cont}
\end{eqnarray}
with $\mathcal{U}_a=\text{diag}[1, e^{\gamma},e^{2\gamma},\cdots,e^{(L_y-1)\gamma}]$ being the amplification matrix and $\mathcal{U}_l=\text{diag}[1, e^{-\gamma},e^{-2\gamma},\cdots,e^{-(L_y-1)\gamma}]$ being the loss matrix (assuming $\gamma >0$).
In a similar fashion, $\mathfrak{t}'_{\text{h}}(\gamma)=\mathcal{U}_l\mathfrak{t}'_{\text{h}}(0)\mathcal{U}_a$.
Thus, more amplifications occur when the chiral edge state has a large width~($\alpha$ closer to $\pi/4$ or $3\pi/4$) than when it has a small width~($\alpha$ closer to $\pi/2$ or $0$).
We emphasize that at the decoupled point of $\alpha=\pi/2$, the chiral edge states have a vanishing wave-packet width, which leads to a quantized transmission $1$.

As a further check that this is indeed a contact effect, we show how the transmission depends on the length of the system, $L_x$, as well as on its width, $L_y$, [see Fig.~\ref{fig: horizontal_transmission}(b)]. 
In the topological phase, only the edge modes contribute to transmission, and since they do not backscatter we indeed observe that the transmission is independent of $L_x$ (apart from very small $L_x$, which is due to finite-size effects).
However, the transmission increases monotonically with the system width, $L_y$, even for system widths which are much larger than the width of the edge states.
This is because changing the width of the system also means changing the width of the two leads (or contacts).

\begin{figure}
\centering
\includegraphics[width=\columnwidth]{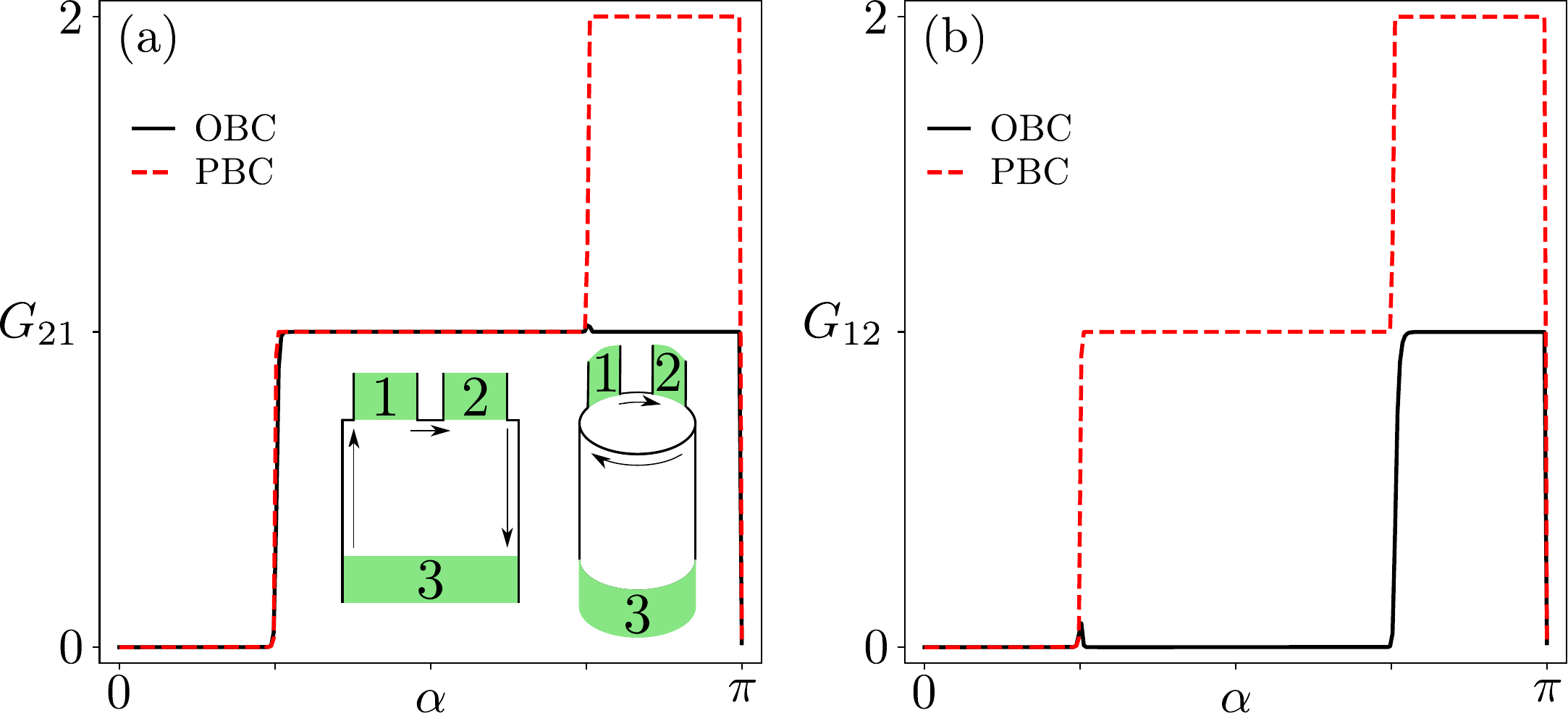}
\caption{The transmission between top two leads as a function of $\alpha$. $G_{ij}$ corresponds to the transmission from $j$ to $i$. 
All plots are with $\gamma=0.3$, and the top two leads each have a width of $20$ unit cells. 
The total system size $75\times 75$ unit cells. 
The inset is a sketch the three terminal geometry for both OBC and PBC.
\label{fig: three_terminal}}
\end{figure}

Finally, we show that the contact effect disappears when changing the lead orientation. 
We attach two vertical leads (labeled as $1$ and $2$) to the top edge and one wide vertical lead (labeled as $3$) to the bottom edge, while in the horizontal direction we consider both OBC as well as periodic boundary conditions (PBC), as shown in Fig.~\ref{fig: three_terminal}.
Here, the lead $3$ is used to eliminate the influence of the bottom boundary mode, such that the transmission between lead $1$ (top left) and lead $2$ (top right) is only due to the top boundary mode.
However, now there is no amplification and loss process described by Eq.\ \eqref{eq:cont}.
Thus, the transmission due to the chiral edge mode is quantized (see Fig.~\ref{fig: three_terminal}). 
With OBC, the unidirectional mode contributes to transmission only from lead 1 to lead 2 ($G_{21}$), while the transmission in the opposite direction ($G_{12}$) vanishes, as expected. 
With PBC in the horizontal direction, however, $G_{12}=G_{21}=1$, since the chiral edge state is now allowed to propagate also across the periodic boundary.

Due to the counter-propagating nature of boundary modes in the weak topological phase ($|\alpha|>3\pi/4$), in the OBC system there exists a transmitting channel from lead $1$ to lead $2$ and vice versa. 
This leads to a quantized transmission with OBC,  $G_{12}=G_{21}=1$ in the weak topological phase.
With PBC in the horizontal direction, we instead observe $G_{12}=G_{21}=2$, consistent with the presence of an extra transmitting channel that connects the leads 1 and 2 across the periodic boundary.

\subsection{Numerical instability}
\label{sec:ni}

\begin{figure}
\centering
\includegraphics[width=0.7\columnwidth]{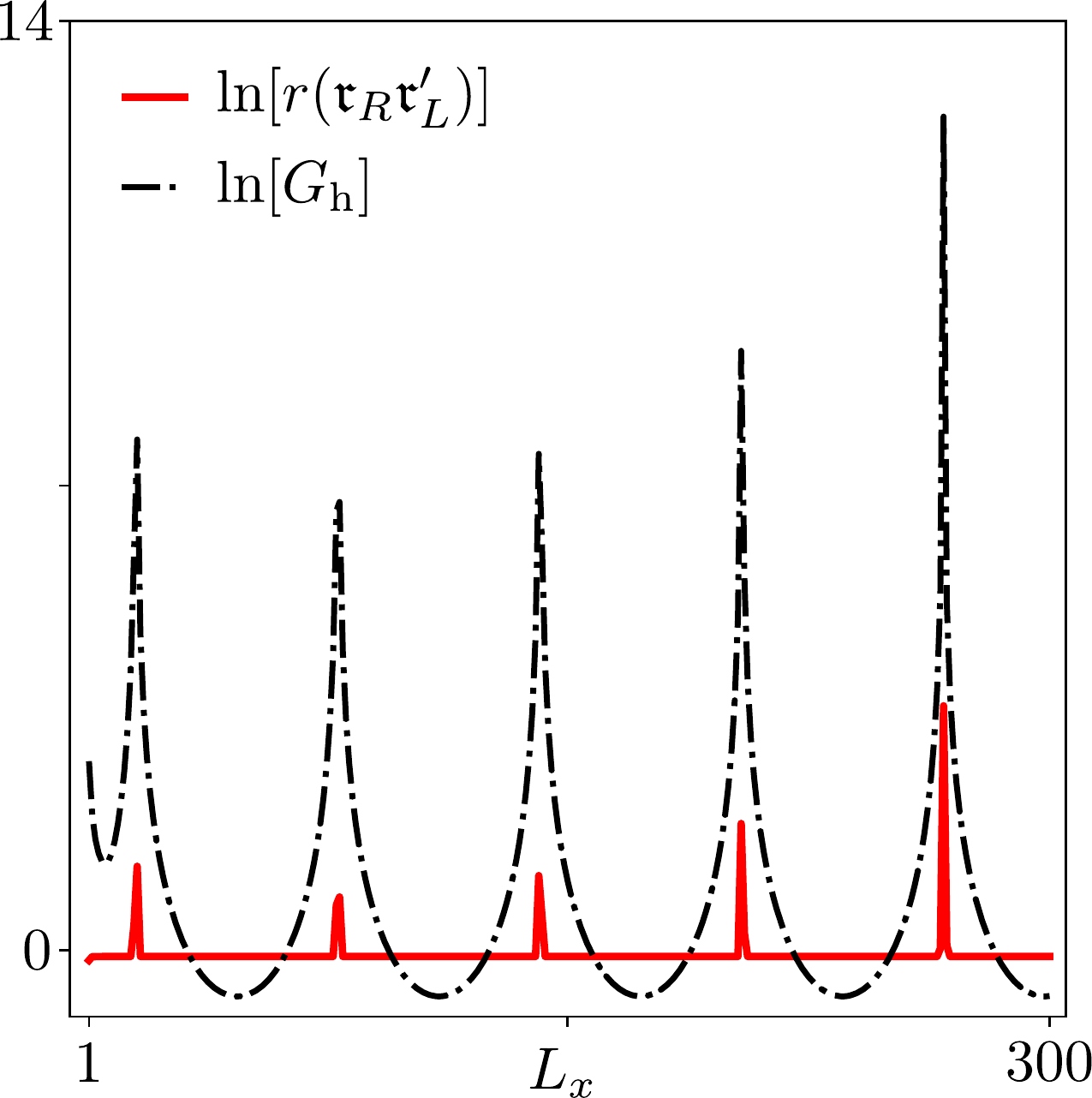}
\caption{$\text{ln}[G_{\text{h}}]$ and $\text{ln}[r(\mathfrak{r}_R\mathfrak{r}'_L)]$ as a function of system size $L_x$. 
All plots are for $L_y=50$, $\alpha=\pi/4$, and $\gamma=0.05$.
\label{fig: divergece}}
\end{figure}

When PBC are applied in the $y$-direction, the horizontal transmission $G_{\text{h}}$ would be divergent due to the singular behaviors of EPs, 
\begin{eqnarray}
\lim_{\textbf{k}\rightarrow\textbf{k}_{\text{EP}}}\text{Im}(\varepsilon)\rightarrow
  0,\quad \lim_{\textbf{k}\rightarrow\textbf{k}_{\text{EP}}}\frac{\partial \text{Re}(\varepsilon)}{\partial k_x}\rightarrow\infty.
\end{eqnarray}
Such an infinite group velocity implies a numerical instability appearing in the calculation of the Landauer-B{\"u}ttiker formula, which corresponds to an ill-defined scattering matrix. 
This instability can be understood when expressing the overall scattering matrix $S_\text{h}$ of the network model by a Redheffer star product \cite{Redheffer}
\begin{align}\label{eq:combiningS}
\begin{split}
& \mathfrak{r}_\text{h} = \mathfrak{r}_L + \mathfrak{t}_L' \mathfrak{r}_R (1 - \mathfrak{r}_L' \mathfrak{r}_R)^{-1} \mathfrak{t}_L, \\
& \mathfrak{t}_\text{h} = \mathfrak{t}_R (1 - \mathfrak{r}_L' \mathfrak{r}_R)^{-1} \mathfrak{t}_L, \\
& \mathfrak{t}'_\text{h} = \mathfrak{t}_L' (1 - \mathfrak{r}_R \mathfrak{r}_L')^{-1} \mathfrak{t}_R', \\
& \mathfrak{r}'_\text{h} = \mathfrak{r}_R' + \mathfrak{t}_R \mathfrak{r}_L' (1 - \mathfrak{r}_R \mathfrak{r}_L')^{-1} \mathfrak{t}_R'.
\end{split}
\end{align}
Here $\mathfrak{r}^{(\prime)}_{L,R}$ and $\mathfrak{t}^{(\prime)}_{L,R}$ are the reflection and transmission matrix of the left and right halves of the network model.
The inverse matrix $(1 - \mathfrak{r}_L' \mathfrak{r}_R)^{-1}$ represents an infinite series of backscattering process between two halves, given by
\begin{eqnarray}
\sum_{i=0}^{\infty}(\mathfrak{r}_L' \mathfrak{r}_R)^i=(1 - \mathfrak{r}_L' \mathfrak{r}_R)^{-1},\label{eq:nuemann}
\end{eqnarray}
where the spectral radius $r(\mathfrak{r}_L' \mathfrak{r}_R)$ is always smaller than $1$ in the unitary limit.
However, the existence of the non-Hermitian terms allows $r(\mathfrak{r}_L' \mathfrak{r}_R)$ to be larger than $1$, due to the amplification loops introduced by the $e^\gamma$ terms.
Thus, the left-hand side of Eq.~\eqref{eq:nuemann} becomes a divergent matrix series, which causes $G_{\text{h}}$ under PBC to be numerically unstable~[see Fig.~\ref{fig: divergece}].

\section{Transport properties of the Dirac cone and bulk Fermi arc}
\label{sec:transport_clean}

\begin{table}[h!]
\begin{center}
\caption{Finite size scaling fit for the bulk transmission $G_{\text{v}}^{\text{t}\rightarrow\text{b}}$ of the clean system. 
The size ratio is $L_x/L_y=10$ and $L_x\in [60, 120]$.
}
\label{table:clean}
\begin{tabular}{|l|c|c|c|c|r}
\hline
\text{parameter} & $\nu$ & $\gamma_c$ & $\gamma_{\text{EP}}$ & $\text{ln}(G_{\text{v}}^{\text{t}\rightarrow\text{b}})_c$\\
\hline
$\alpha = 0.15\pi$ & $1$ & $0.937$ & $0.937$ & $1.296(1)$\\
\hline
$\alpha=0.25\pi$ & $1$ & $0$ & $0$ & $0$\\
\hline
$\alpha=0.3\pi$ & $1$  & $0.4499$ & $0.4499$ & $1.384$\\
\hline
$\alpha=0.375\pi$ & $1$ & $1.212$ & $1.212$ & $1.220(1)$\\
\hline
\end{tabular}
\end{center}
\end{table}

Here we turn to the vertical transmission, $G_{\text{v}}$.
Since there is no skin effect when open boundaries are introduced along the horizontal direction, the system shows a similar ribbon spectrum both in OBC and PBC, except of course for the additional chiral edge states occurring with OBC.
Near the EPs, both $\partial \text{Re}(\varepsilon)/\partial k_y$ and $\text{Im}(\varepsilon)$ are finite and thus there are no numerical instabilities.
Although the vertical transmission $G_{\text{v}}$ comes with no infinite amplification loops, a directional amplification~(suppression) from top~(bottom) to bottom~(top) implies $G_{\text{v}}^{\text{t}\rightarrow\text{b}}\neq G_{\text{v}}^{\text{b}\rightarrow\text{t}}$.

To understand the behavior of the vertical transmission $G_{\text{v}}$, we first consider the unitary limit, $\gamma=0$, in which the Dirac points occur at $\alpha=\pm\pi/4$ and $\pm3\pi/4$ and have a quantized transmission $1$. 
As $\gamma$ is increased, however, the network enters non-unitary regimes and is characterized by gapless phases containing EPs  and the bulk Fermi arc. 
The existence of the EPs and of the bulk Fermi arc $\text{Re}(\varepsilon)=0$ leads to a non-zero $G_\text{v}$. 

\begin{figure}
\centering
\includegraphics[width=1\columnwidth]{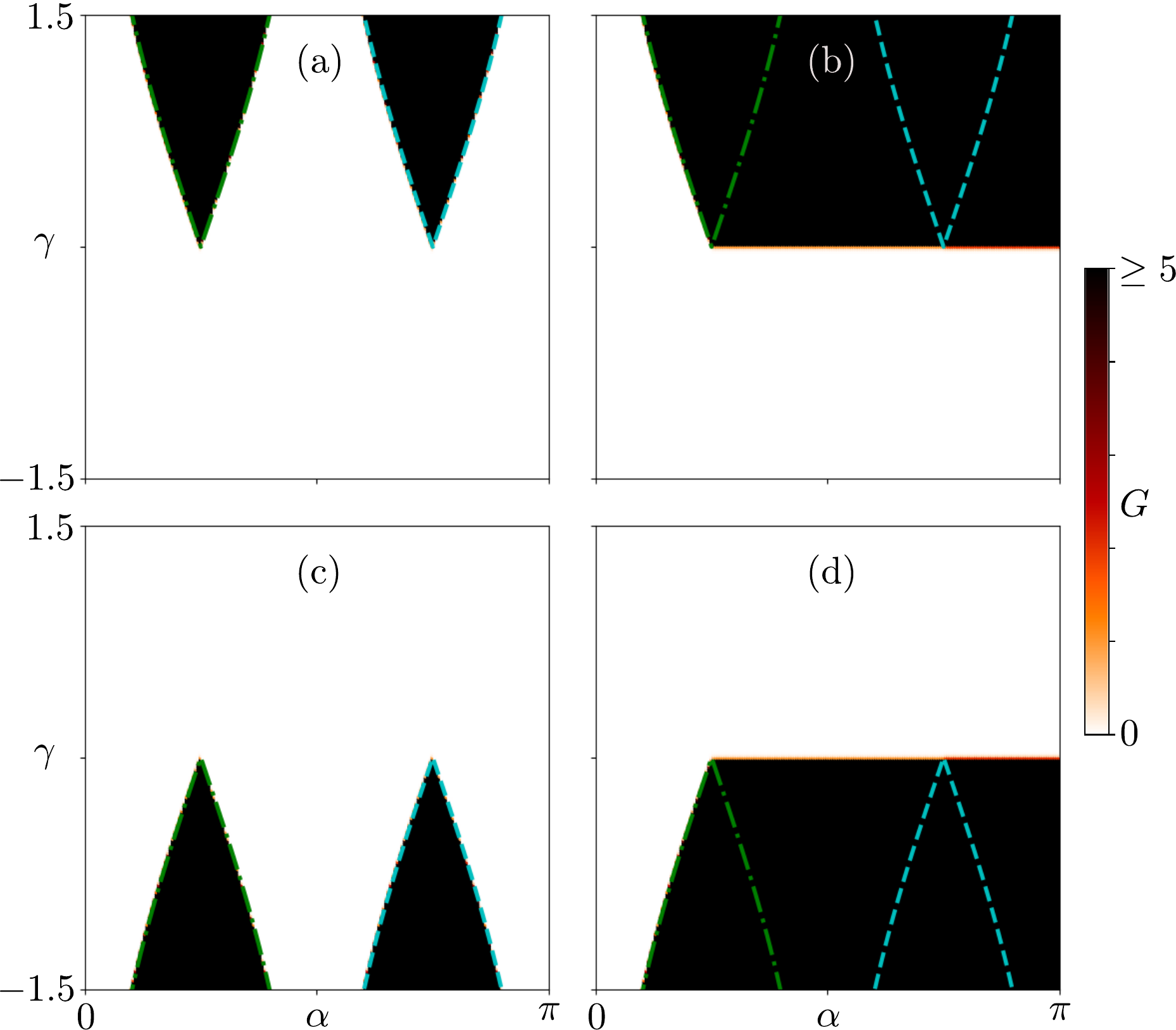}
\caption{The vertical transmission as a function of $\alpha$ and $\gamma$ both with periodic boundary condition (a, c) and with open boundary condition (b, d). 
(a-b) represent the transmission from top to bottom, (c-d) represent the transmission from bottom to top. 
All plots are obtained for a system size $60\times 60$ unit cells.
\label{fig:vertical_transmission}}
\end{figure}

In Fig.~\ref{fig:vertical_transmission} we show the vertical transmission as a function of $\alpha$ and $\gamma$, for both PBC as well as OBC.
The green dashed-dotted line and the cyan dashed line determined by Eqs.~\eqref{eq:boundary_1} and \eqref{eq:boundary_2} indicate the Dirac points appearing at $\textbf{k} = (0,\pi)$ and $(\pi, 0)$, respectively.
Between these lines, we observe a large transmission characterizing the gapless phase with EPs. 
In addition, the OBC plots show transmitting regions which do not appear in the PBC plots.
These correspond to nonzero transmission through the topological edge modes of the system.
Notice that the transmission plots are asymmetric with respect to changing the sign of $\gamma$, which is consistent with the directional nature of amplification and attenuation, $G_{\text{v}}^{\text{t}\rightarrow\text{b}}\neq G_{\text{v}}^{\text{b}\rightarrow\text{t}}$.

\begin{figure}
\centering
\includegraphics[width=1\columnwidth]{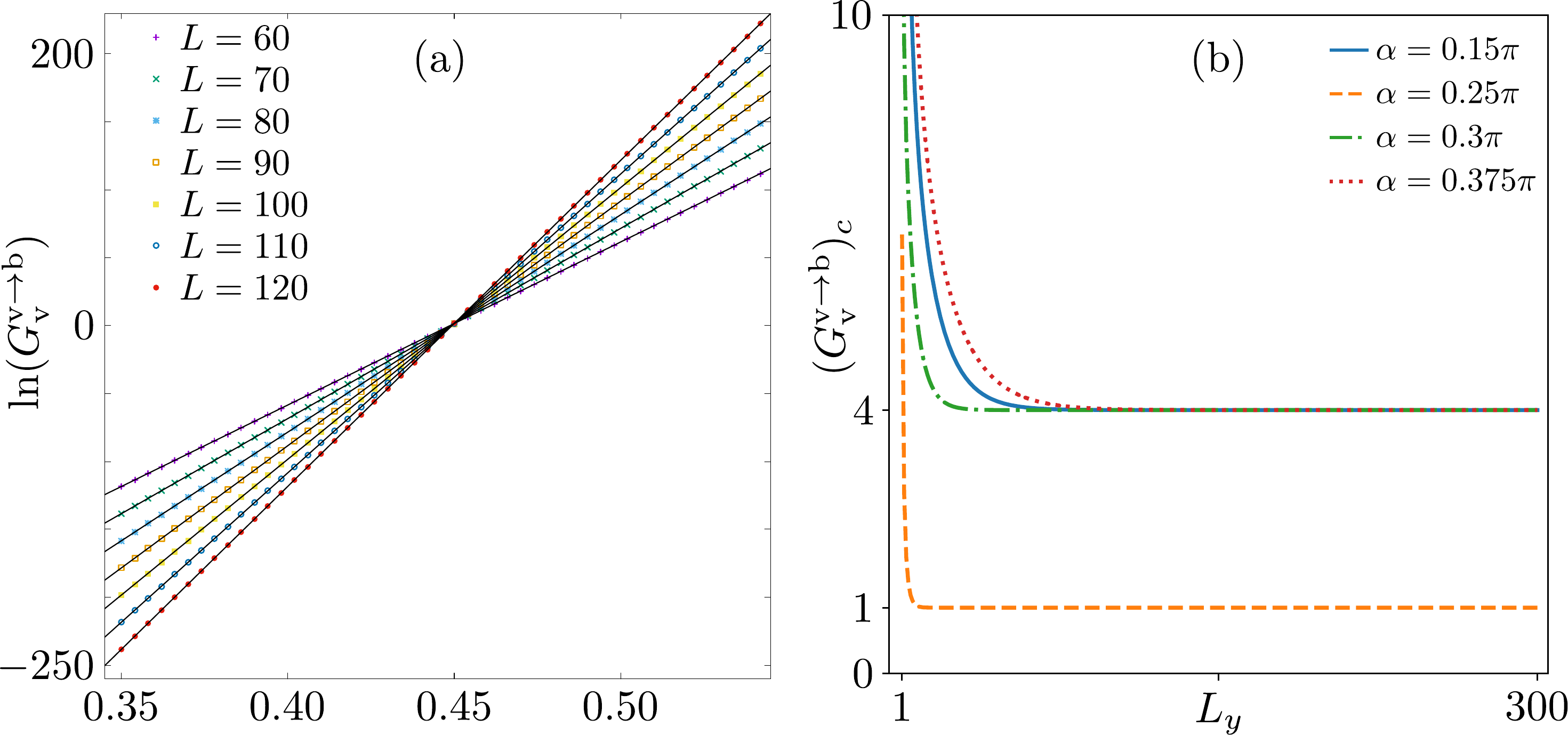}
\caption{(a) The logarithm of transmission $G_{\text{v}}^{\text{t}\rightarrow\text{b}}$ of different system sizes with periodic boundary conditions as a function of $\gamma$ for $\alpha=0.3\pi$ and system size $10L\times L$ ($L_x\times L_y$). 
The solid lines are the fit using Eq.~\eqref{eq:clean_fit}.
(b) The critical transmission for different $\alpha$ as a function of system size $L_y$. 
This plot is obtained for $L_x=20$. When $\alpha=\pi/4$, the Dirac point occurs for $\gamma=0$, and thus shows the critical transmission characteristic of Hermitian Dirac cones, $G=1$. 
For all other values of $\alpha$ the transmission is 4. 
\label{fig: scaling_fit_clean}}
\end{figure}

When the system transitions from a gapped phase to a gapless phase characterized by EPs (green and cyan lines in Fig.~\ref{fig:vertical_transmission}), we notice that the transmission changes from decreasing exponentially with system size to increasing exponentially with system size.
We apply a finite size scaling analysis to $G_{\text{v}}^{\text{t}\rightarrow\text{b}}$ with PBC and $L_y/L_x=10$. 
Our code, numerical data, and the scripts used for fitting are available as part of the Supplemental Material.
Denoting the system size at fixed aspect ratio by $L$, we have
\begin{eqnarray}\label{eq:clean_fit}
\text{ln}G_{\text{v}}^{\text{t}\rightarrow\text{b}}\approx f_0+b_0(\gamma-\gamma_c)L^{1/\nu}.
\end{eqnarray}
We obtain $b_0\approx20$, $\nu=1$, and $\gamma_c=\gamma_{\text{EP}}$ even for different $\alpha$ (see Table \ref{table:clean}). 
Here $\gamma_{\text{EP}}$ corresponds to the analytically determined value at which the Dirac points appear, meaning the point at which the EPs are overlapping in momentum space, just after they were created pairwise, or just before annihilating pairwise.
For instance, in Fig. \ref{fig: scaling_fit_clean}(a), we find that below $\gamma_c\simeq 0.4499$, $G_{\text{v}}^{\text{t}\rightarrow\text{b}}$ decreases exponentially with increasing $L$, characteristic of a gapped phase.
Above $\gamma_c$, $G_{\text{v}}^{\text{t}\rightarrow\text{b}}\propto e^{2L}$.
This is in contrast to the behavior characteristic of Hermitian gapless systems, in which the transmission grows at most linearly with system size (ballistic regime).

The scattering matrix characterizing the finite-sized network model with leads attached to its top and bottom boundaries can be expressed as
\begin{eqnarray}
S_{\text{v}}(\gamma)=\begin{pmatrix}
\mathfrak{r}_{\text{v}}&e^{-\gamma L_y}\mathfrak{t}'_{\text{v}}\\e^{\gamma L_y}\mathfrak{t}_{\text{v}} & \mathfrak{r}'_{\text{v}}
\end{pmatrix}.\label{eq:final_s}
\end{eqnarray}
Here $S_{\text{v}}(\gamma=0)$ represents the scattering matrix of the unitary CC network. 
For gapped phases, we  have $\text{tr}(\mathfrak{t}\mathfrak{t}^\dagger)=\text{tr}(\mathfrak{t}'\mathfrak{t}'^\dagger)\propto e^{-\Delta L_y}$, with $\Delta$ proportional to the bulk gap of the Hermitian model.
Since the gain~(loss) introduces an amplification (a reduction)
$e^{2\gamma L_y}$ ($e^{-2\gamma L_y}$) to $G_{\text{v}}^{\text{t}\rightarrow\text{b}}$ ($G_{\text{v}}^{\text{b}\rightarrow\text{t}}$), the gapped-gapless transition occurs due to the competition between $\Delta$ and $\gamma$
\begin{eqnarray}
G_{\text{v}}^{\text{t}\rightarrow\text{b}}\propto e^{-(\Delta -2\gamma)L_y}.
\end{eqnarray}
Thus, if $\gamma>\Delta/2$, the system enters a gapless phase.

In the unitary limit the Dirac point has a quantized transmission $G=1$.
Quite surprisingly, we find that in the non-unitary network this is no longer the case (see Table~\ref{table:clean}).
Instead, there is a new critical transmission
\begin{eqnarray}
  (G_{\text{v}}^{\text{t}\rightarrow\text{b}})_c=4,
  \quad
  (G_{\text{v}}^{\text{b}\rightarrow\text{t}})_c=0,\label{eq:critical_transmission}
\end{eqnarray}
as shown in Fig.~\ref{fig: scaling_fit_clean}(b).
These values can be understood analytically from the behavior of slices having $L_x\to\infty$, such that $k_x$ is a good quantum number (see Appendix \ref{app: sca_dirac}). 
Note that in Fig.~\ref{fig: scaling_fit_clean}(b) the quantized transmission is reached in the limit of large system sizes.
For the smaller network models used in the scaling fit of Fig.~\ref{fig: scaling_fit_clean}(a), finite-size effects prevent the observation of $(G_{\text{v}}^{\text{t}\rightarrow\text{b}})_c=4$, meaning ${\rm ln} (G_{\text{v}}^{\text{t}\rightarrow\text{b}})_c\simeq 1.386$, as can be seen in Table~\ref{table:clean}.
The closest fit value is 1.384, obtained for $\alpha=0.3\pi$, which shows smaller finite-size effects than $\alpha=0.15\pi$ or $0.375\pi$, as can be seen in Fig.~\ref{fig: scaling_fit_clean}(b).

\section{Disorder and critical exponent}
\label{sec:disorder}

We now turn to the effects of disorder.
The latter is modeled as a random variation of the mixing angle, $\alpha \to \alpha + \delta\alpha$, with $\delta\alpha$ drawn independently for each node of the network model from the uniform distribution $[-W, W]$, with $W$ the disorder strength.
Due to the numerical instability discussed earlier, we focus mainly on the vertical transmission for a network with PBC.
For the same reason, we keep $\gamma$ constant and only consider disorder in $\alpha$.

In the unitary CC network model, the random angle disorder $\delta\alpha$ preserves PHS and leads to an insulator-metal transition~\cite{Chalker_tqhe, Evers_rmp}.
The system supports a trivial phase, a strong topological phase, and a weak topological phase with weak disorder.
With strong disorder, a metallic phase appears for all $\alpha$.
At the phase boundaries, the localization length diverges as $\xi\approx |x-x_c|^{-\nu}$.
Here $x$ is a control parameter, such as $\alpha$ or the disorder strength $W$, $x_c$ is the critical point of $x$, and $\nu$ is the critical exponent, which is equal to $1$ for AZ class D systems in two dimensions. 

\begin{figure}
\centering
\includegraphics[width=1\columnwidth]{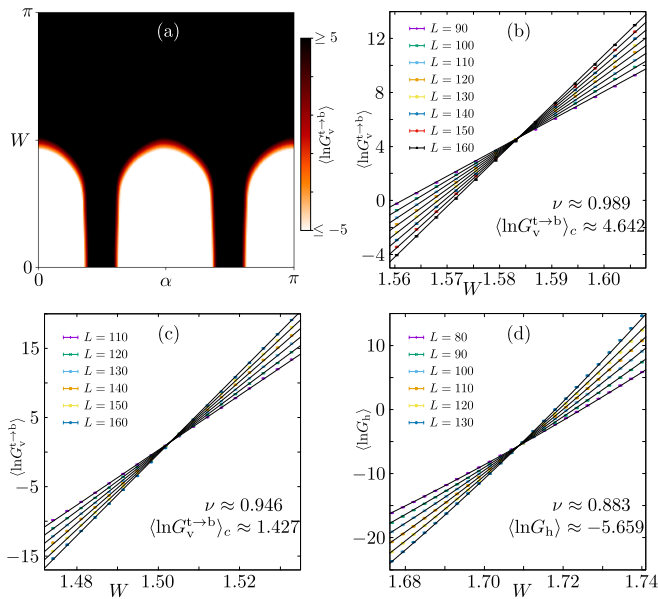}
\caption{(a) The average transmission $\langle\text{ln} G_{\text{v}}^{\text{t}\rightarrow\text{b}}\rangle$ as a function of disorder strength $W$ and $\alpha$. 
The system size is $30\times 30$ and $\gamma=0.5$. (b) and (c) are the finite size scaling fit of $\langle \text{ln}G_{\text{v}}^{\text{t}\rightarrow\text{b}}\rangle$ for $(\gamma, \alpha)=(0.5,\pi/2)$ and $(0.2,0.35\pi)$, respectively. 
(d) the finite size scaling fit of $\langle\text{ln} G_{\text{h}}\rangle$ with open boundary condition and $(\gamma,\alpha)=(0.5,0)$. 
\label{fig: pd_disorder}}
\end{figure}

Since the non-unitary network has a bulk topological semimetal phase in the clean limit, its disordered phase diagram also supports phase transitions from the delocalized phase to an insulating one in the weak disorder limit~[see Fig.\ \ref{fig: pd_disorder}(a)].
This is in contrast to the unitary Cho-Fisher model, in which only direct, insulator-to-insulator transitions are present for small disorder strengths.
In the strong disorder regime, however, all the phases evolve into a delocalized phase, similar to the Cho-Fisher model.
Here, however, the transmission increases exponentially with system size, as can also be deduced from Eq.~\eqref{eq:final_s}, which is also valid for a disordered system.
This is in contrast to the Hermitian system, in which the transmission grows logarithmically with system size in the metallic phase.
For this reason, we dub the delocalized phase of the non-Hermitian network a `supermetal'.

\begin{table*}[tb]
\begin{center}
\caption{Fit results for the vertical bulk transmission of the disorder system with a system size $L\times L$. 
$N$ is the number of degrees of freedom used in the fit, GOF is the goodness of fit, and reduced $\chi^2$ is the variance of residuals. 
The numerics has been averaged over more than $7000$ disorder realizations.}
\label{table: finite_disorder_fit_vertical}
\begin{tabular}{|c|c|c|c|c|c|c|c|c|c|c|}
\hline
$(\alpha,\gamma)$ & $(n,m_0,q_r)$ & $L$ & $W$ & $W_c$ & $\nu$ & $\langle\text{ln}G_{\text{v}}^{\text{t}\rightarrow \text{b}}\rangle_c$ &
$\chi^2$ & $N$ & \text{GOF} \\
\hline
\multirow{2}{1.55cm}{$(0.4\pi,0.5)$} & \multirow{2}{1cm}{$(0,1,2)$} & $[110, 160]$ & $[1.436, 1.478]$ & $1.4566~[1.4562,1.4568]$ & $0.979~[0.955,1.006]$ & $4.707~[4.581,4.785]$ & $1.027$ & $67$ & $0.91$ \\
&& $[110, 160]$ & $[1.412, 1.5]$ & $1.4566~[1.4563,1.4568]$ & $0.974~[0.966,0.984]$ & $4.707~[4.629,4.763]$ & $1.041$ & $151$ & $0.80$ \\
\hline
\multirow{2}{1.55cm}{$(0.5\pi,0.5)$} & \multirow{2}{1cm}{$(0,1,1)$} & $[110, 160]$ &$[1.555,1.612]$ & $1.5840~[1.5837,1.5844]$ & $0.990~[0.969,1.013]$ & $4.796~[4.689,4.910]$ & $0.962$ & $86$ & $0.84$\\
&& $[90, 160]$ &$[1.559,1.608]$ & $1.5836~[1.5834,1.5839]$ & $0.989~[0.973,1.007]$ & $4.642~[4.575,4.726]$ & $0.994$ & $100$ & $0.97$\\
\hline
\multirow{2}{1.55cm}{$(0.35\pi,0.2)$} & \multirow{2}{1cm}{$(0,1,1)$}&$[110,160]$&$[1.473,1.534]$&$1.5029~[1.5025,1.5032]$ & $0.946~[0.926,0.966]$& $1.427~[1.262,1.573]$ &$1.046$&$104$&$0.82$\\
&&$[80,160]$&$[1.483,1.525]$&$1.5026~[1.5024,1.5028]$ & $0.928~[0.910,0.947]$& $1.246~[1.155,1.358]$ &$0.980$&$104$&$0.92$\\
\hline
\end{tabular}
\end{center}
\end{table*}

\begin{table*}[tb]
\begin{center}
\caption{Fit results for the vertical bulk transmission of the disorder system with a system size $L\times 4L$. 
$N$ is the number of degrees of freedom used in the fit. 
The numerics has been averaged over $2000$ disorder realizations.}
\label{table: finite_disorder_fit_vertical_ratio}
\begin{tabular}{|c|c|c|c|c|c|c|c|c|c|c|}
\hline
$(\alpha,\gamma)$ & $(n,m_0,q_r)$ & $L$ & $W$ & $W_c$ & $\nu$ & $\langle\text{ln}G_{\text{v}}^{\text{t}\rightarrow \text{b}}\rangle_c$ &
$\chi^2$ & $N$ & \text{GOF} \\
\hline
\multirow{2}{1.55cm}{$(0.5\pi,0.5)$} & $(0,1,2)$ & $[80, 160]$ &$[1.545,1.636]$ & $1.5918~[1.5916,1.5919]$ & $0.988~[0.980,0.995]$ & $0.932~[0.716,1.112]$ & $0.987$ & $112$ & $0.93$\\
&$(0,2,1)$& $[80, 160]$ &$[1.55,1.631]$ & $1.5920~[1.5917,1.5922]$ & $0.985~[0.974,0.998]$ & $1.224~[0.944,1.410]$ & $1.014$ & $94$ & $0.93$\\
\hline
\multirow{2}{1.55cm}{$(0.35\pi,0.2)$} &$(0,1,2)$&$[90,160]$&$[1.46,1.553]$&$1.5078~[1.5076,1.5081]$ & $0.946~[0.935,0.956]$& $-9.245~[-9.722,-8.776]$ &$1.03$&$107$&$0.84$\\
&$(0,2,1)$&$[90,160]$&$[1.46,1.553]$&$1.5082~[1.5079,1.5084]$ & $0.944~[0.932,0.954]$& $-8.531~[-8.992,-8.113]$ &$0.969$&$107$&$0.86$\\
\hline
\end{tabular}
\end{center}
\end{table*}

\begin{table*}[tb]
\begin{center}
\caption{Fit results for the horizontal transmission of the disorder open boundary system with a system size $L\times L$. 
$N$ is the number of degrees of freedom used in the fit. 
The numerics has been averaged over $10000$ disorder realizations.}
\label{table: finite_disorder_fit_horizontal}
\begin{tabular}{|c|c|c|c|c|c|c|c|c|c|c|}
\hline
$(\alpha,\gamma)$& $(n,m_0,q_r)$ & $L$ & $W$ & $W_c$ & $\nu$ & $\langle\text{ln}G_{\text{h}}\rangle_c$ & $\chi^2$ & $N$ & \text{GOF} \\
\hline
\multirow{2}{1.cm}{$(0,0.5)$} & $(0,1,1)$ & $[80, 130]$ & $[1.695, 1.722]$ & $1.7089~[1.7085,1.7092]$ & $0.889~[0.839,0.941]$ & $-5.547~[-5.709,-5.382]$ & $1.009$ & $56$ & $0.98$ \\
&  $(0,1,2)$  & $[80, 130]$ & $[1.677, 1.74]$ & $1.7087~[1.7086,1.7090]$ & $0.883~[0.869,0.901]$ &  $-5.659~[-5.753,-5.545]$ & $0.965$ & $133$ & $0.93$ \\
\hline
\end{tabular}
\end{center}
\end{table*}

Similar to the results of Ref.~\cite{Kawabata_prl2021}, we find that the transmission of the network model varies exponentially for different disorder configurations, leading to an exponentially broad distribution of transmissions, which is poorly characterized by its average.
We circumvent this problem by focusing instead on the average of the `typical transmission', $\langle\text{ln}G_{\text{v}}^{\text{t}\rightarrow\text{b}}\rangle$, which we use as a scaling variable in the following.
To obtain the critical exponent $\nu$, we apply a finite-size scaling fit to $\langle\text{ln}G_{\text{v}}^{\text{t}\rightarrow\text{b}}\rangle$, at fixed aspect ratio 1, with
\begin{eqnarray}
\langle\text{ln}G_{\text{v}}^{\text{t}\rightarrow \text{b}}\rangle &=&F(u_0L^{1/\nu},u_1L^{y})\nonumber\\
&=&\sum_{k=0}^{n}(u_1L^{y})^k\sum_{j=0}^{m_k}(u_0L^{1/v})^jF_{kj},\label{eq:fit_equation}
\end{eqnarray}
where $u_1L^y$ represents the contribution of the irrelevant exponent $y<0$, and goes to $0$ if $L\rightarrow \infty$. $u_0L^{1/v}$ represents the contribution of the relevant exponent, $\nu$. 
Here $u_0=\sum_{i=1}^{q_r}a_i(x-x_c)^i$ and $u_1=\sum_{p=0}^{q_i}b_p(x-x_c)^p$.
In Tables \ref{table: finite_disorder_fit_vertical} and \ref{table: finite_disorder_fit_vertical_ratio}, and Fig.~\ref{fig: pd_disorder}(b-c), for different $\alpha$ and $\gamma$, $\nu$ is always close to $1$, which is consistent with the case of the AZ class D system. 
However, the critical transmission appears to no longer have a universal value.
Given the system sizes we can reach, we believe that this is a consequence of finite-size effects.

We provide a heuristic justification for having a critical exponent $\nu=1$, as well as a critical transmission which should be universal.
In the presence of disorder in the mixing angle $\alpha$, Eq.~\eqref{eq:final_s} is still valid.
When the unitary system is in the insulating phase with $\langle G\rangle\propto
e^{-L/\xi(\alpha,W)}$ (far away from the critical point), the non-unitary system
has $\text{ln}\langle G_{\text{v}}^{\text{t}\rightarrow \text{b}}\rangle = \mathrm{const.} + (2\gamma -1/\xi(\alpha,W))L$, which suggests a localization-delocalization transition at $\xi(\alpha,W)=1/2\gamma$ with $\nu=1$.
When the unitary network is in an insulating phase but close to the critical point, ideally the average transmission would be $\langle G\rangle=G_c+(W-W_c)L^{1/\nu}+\mathcal{O}^2(W-W_c),$ where $\mathcal{O}^2(W-W_c)$ is the second-order infinitesimal and $G_c$ is the universal critical transmission of two-dimensional class D systems. 
On the other hand, the transmission of the non-unitary system would be $\langle G_{\text{v}}^{\text{t}\rightarrow \text{b}}\rangle\approx (G_c+(W-W_c)L^{1/\nu})e^{2\gamma L}$, which can be expanded as
\begin{eqnarray}
\text{ln}\langle G_{\text{v}}^{\text{t}\rightarrow \text{b}}\rangle \approx  2\gamma L + \text{ln}G_c + \frac{W-W_c}{G_c}L^{1/\nu} +\cdots.\label{eq:log_average}
\end{eqnarray}
Since Hermitian class D systems have $\nu=1$ in two dimensions, the above equation becomes
\begin{eqnarray}\label{eq:lavgfinal}
\text{ln}\langle G_{\text{v}}^{\text{t}\rightarrow \text{b}}\rangle \approx \text{ln}G_c + \frac{W- (W_c - 2\gamma G_c)}{G_c}L,
\end{eqnarray} 
where $\mathcal{O}^2(\frac{W-W_c}{G_c})$ is neglected.
According to Eq.~\eqref{eq:fit_equation}, it corresponds to a linear dependence on $L$, implying a universal exponent $\nu=1$ also in the non-Hermitian case.
This is consistent with our fit results.

Equation \eqref{eq:lavgfinal} suggests that the effect of a nonzero $\gamma$ is to renormalize the value of the critical disorder strength as $W_c \to W_c - 2\gamma G_c$. 
This behavior also matches our numerical observations: increasing $\gamma$ leads to a decrease of the critical disorder strength at which the insulator-supermetal transition happens. 
Finally, Eq.~\eqref{eq:lavgfinal} indicates that the critical transmission takes a universal value equal to that of Hermitian systems, $G_c$. 
This is not the case for the scaling analysis presented in Tables \ref{table: finite_disorder_fit_vertical} and \ref{table: finite_disorder_fit_vertical_ratio}, suggesting that the fitted values of 
$\langle\text{ln}G_{\text{v}}^{\text{t}\rightarrow \text{b}}\rangle_c$ are influenced by the finite size of the systems we are able to simulate.
Note, however, that the above discussion is only qualitative in nature, since we
have used the logarithm of the average, $\text{ln}\langle G_{\text{v}}^{\text{t}\rightarrow \text{b}}\rangle$, and not the average of the logarithm, $\langle\text{ln}G_{\text{v}}^{\text{t}\rightarrow \text{b}}\rangle$, as in the numerical fits.

As for the horizontal direction, due to the numerical instability, we can only apply a finite size scaling fit to $\langle\text{ln}G_{\text{h}}\rangle$ with OBC, which induces an amount of finite-size effects automatically.
In Table \ref{table: finite_disorder_fit_horizontal}, the fit results also give a critical exponent $\nu\approx 0.89$, which seems close to the value $1$.

\section{Conclusion and outlook}
\label{sec:conclusion}

We have shown that a network model can exhibit both exceptional points as well as the non-Hermitian skin effect, which lead to several new phenomena.
On the boundary, the non-Hermitian skin effect results in the non-Hermitian contact effect, which destroys the quantized conductance of the chiral edge states.
In the bulk, the existence of exceptional points dominates the transport properties, which show a numerical instability when the system can feel the branch point singularity ($\partial_{k_x}\text{Re}(\varepsilon)\rightarrow \infty$). 
In contrast, $\partial_{k_y}\text{Re}(\varepsilon)$ is finite in the perpendicular direction. 
However, the transmission still experiences a directional amplification~(or loss).
This amplification results in an unconventional quantized transmission $G=4$ for the Dirac points, and drives the system into a gapless phase.
Such a gapless phase reshapes the disordered phase diagram of systems with PHS$^\dagger$, leading to the formation of a supermetal in which the transmission grows exponentially with system size.
Finally, the finite size scaling analysis provides a critical exponent $\nu\approx 1$, which is same as the critical exponent of the AZ class D system.
However, the critical transmission is no longer universal, at least as far as we are able to determine, which is most probably due to finite-size effects.
We provide a heuristic argument for why this might be so, based on an analytic connection between the scattering matrix of the unitary and non-unitary network models.

Our work opens several directions for studying non-Hermitian topology.
As prototypical model, the network studied here is a powerful tool for exploring transport and for revealing its relation to the non-Hermitian features like the exceptional points and skin effect.
In addition, the relation of scattering matrices between the unitary and non-unitary network can also be extended to a variety of symmetry classes, providing a way to study the universal behaviors of different localization-delocalization transitions, such as the quantum spin-Hall transition.
The non-Hermitian network model can further be studied, e.g., in the presence of more generic types of disorder, and by using various techniques known in the context of Hermitian network models.
In Appendix \ref{Path integral approach}, we outline an application of the path integral (supersymmetric field theory) approach to the non-Hermitian network model.

Finally, our work opens the possibility of studying the connection between transport and non-Hermitian phenomena in experimental platforms realizing network models.
In fact, the network model we discussed has already been realized in experiment: a small number of unit cells were produced using microwave circuits in Ref.~\cite{Hwsc_model}.
Beyond that, network models can be realized experimentally also as arrays of coupled ring resonators supporting photonic \cite{Hafezi2011, Afzal2020} or plasmonic \cite{Gao2016, Gao2018} modes. 
These systems allow to tune the properties of the node scattering matrices, for instance by adjusting the separation between adjacent resonators, as well as the eigenphase at which the network model is probed, by changing the wavelength of propagating excitations.
Models obeying particle-hole symmetry (which here means real scattering matrices, up to a global phase factor) have already been experimentally realized \cite{Gao2016, Afzal2020}. 
Further, it has been recently suggested that directional amplification and attenuation can be introduced in coupled ring resonator lattices \cite{Zhu2020, Song2020}.

\emph{Note added:} Ref.~\cite{Luoxunlong2021}, which appeared during the final stages of this work, also considers the critical exponents of non-Hermitian two-dimensional systems. 
They work on a Hamiltonian level, examine other symmetry classes than the one we consider, and propose a link between the critical exponents of Hermitian and non-Hermitian systems which is different from ours.

\begin{acknowledgments}
We thank Ulrike Nitzsche for technical assistance, Chen Wang for sharing their notes with us, as well as Kohei Kawabata for useful discussions.
This work was supported by the Deutsche Forschungsgemeinschaft~(DFG, German
Research Foundation) under Germany's Excellence Strategy through the
W\"{u}rzburg-Dresden Cluster of Excellence on Complexity and Topology in Quantum
Matter -- \emph{ct.qmat} (EXC 2147, project-id 390858490).
S.R. is supported
by the National Science Foundation under award number DMR-2001181, and by a Simons Investigator Grant
from the Simons Foundation (Award Number: 566116).
\end{acknowledgments}

\appendix
\section{Symmetry of a non-unitary operator}
\label{app:sym_eo}

For a non-Hermitian system, the symmetry classification of the Hamiltonian is based on time-reversal symmetry (TRS), particle-hole symmetry (PHS), $\text{TRS}^\dagger$, $\text{PHS}^\dagger$, chiral symmetry (CS), sublattice symmetry (SLS), and pseudo-Hermiticity. 
Then, the Ho-Chalker operator (which behaves like a time evolution operator) $\mathcal{S}\sim e^{-iHt}$ obeys the corresponding symmetry constraints in the following way: 
\begin{enumerate}
  \renewcommand{\labelenumi}{(\roman{enumi})}

\item
  TRS: $\mathcal{T}H^{*}(\textbf{k})\mathcal{T}^{-1}=H(-\textbf{k})$ with $\mathcal{T}\mathcal{T}^{*}=\pm1$, which means
  \begin{eqnarray}
    \mathcal{T}\mathcal{S}^{*}(\textbf{k})\mathcal{T}^{-1}=\mathcal{S}^{-1}(-\textbf{k}).
  \end{eqnarray}

\item
  $\text{TRS}^\dagger$: $\mathcal{T}H^{T}(\textbf{k})\mathcal{T}^{-1}=H(-\textbf{k})$ with $\mathcal{T}\mathcal{T}^{*}=\pm1$, this leads to
  \begin{eqnarray}
    \mathcal{T}\mathcal{S}^{T}(\textbf{k})\mathcal{T}^{-1}=\mathcal{S}(-\textbf{k}).
  \end{eqnarray}

\item
  PHS: $\mathcal{P}H^{T}(\textbf{k})\mathcal{P}^{-1}=-H(-\textbf{k})$ with $\mathcal{P}\mathcal{P}^{*}=\pm1$, which means
  \begin{eqnarray}
    \mathcal{P}\mathcal{S}^{T}(\textbf{k})\mathcal{P}^{-1}=\mathcal{S}^{-1}(-\textbf{k}).
  \end{eqnarray}

\item
  $\text{PHS}^\dagger$: $\mathcal{P}H^{*}(\textbf{k})\mathcal{P}^{-1}=-H(-\textbf{k})$ with $\mathcal{P}\mathcal{P}^{*}=\pm1$, this leads to
  \begin{eqnarray}
    \mathcal{P}\mathcal{S}^{*}(\textbf{k})\mathcal{P}^{-1}=\mathcal{S}(-\textbf{k}).
  \end{eqnarray}

\item
  CS: $\mathcal{C}H^{\dagger}(\textbf{k})\mathcal{C}^{-1}=-H(\textbf{k})$, which means
  \begin{eqnarray}
    \mathcal{C}S^{\dagger}(\textbf{k})\mathcal{C}^{-1}=S(\textbf{k}).
  \end{eqnarray}

\item
  SLS: $\mathcal{L}H(\textbf{k})\mathcal{L}^{-1}=-H(\textbf{k})$, which means
  \begin{eqnarray}
    \mathcal{L}S(\textbf{k})\mathcal{L}^{-1}=S^{-1}(\textbf{k}).
  \end{eqnarray}

\item
  Pseudo-Hermiticity: $\eta H^{\dagger}(\textbf{k}) \eta^{-1} = H(\textbf{k})$, then
  \begin{eqnarray}
    \eta S^\dagger (\textbf{k}) \eta^{-1}=S^{-1}(\textbf{k}).
  \end{eqnarray}

\end{enumerate}

Moreover, their corresponding non-unitary scattering matrix will have the same symmetry constraint.
For a given Hamiltonian $H$, the  Mahaux-Weidenm{\"u}ller formula describes the scattering matrix $S$ as
\begin{eqnarray}
S(E)=\frac{1-i\pi K(E)}{1+i\pi K(E)},~K(E)=W^\dagger \frac{1}{E-H} W.
\end{eqnarray}
Here $W$ is the coupling matrix between the leads and the system, which can be chosen such that it commutes with the symmetry operators. 
Then the symmetry constraints for the corresponding scattering matrix can be derived in the following way:
\begin{enumerate}
  \renewcommand{\labelenumi}{(\roman{enumi})}

\item TRS:
  \begin{eqnarray}
    \mathcal{T}K^{*}(E)\mathcal{T}^{-1}&=&W^{\dagger}\frac{1}{E^{*}-H}W=-K(E^{*}),\\
    \mathcal{T}S^{*}(E)\mathcal{T}^{-1}&=&\frac{1+i\pi K(E^{*})}{1-i\pi K(E^{*})}=S^{-1}(E^{*}).
  \end{eqnarray}
  If $\mathcal{T}=I$, $S(E^{*})S^{*}(E)=I$.

\item TRS$^{\dagger}$:
  \begin{eqnarray}
    \mathcal{T}K^{T}(E)\mathcal{T}^{-1}&=&
                                           W^{\dagger}\frac{1}{E-H}W=K(E),\\
    \mathcal{T}S^{T}(E)\mathcal{T}^{-1}&=&
                                           \frac{1-i\pi K(E)}{1+i\pi K(E)}=S(E).
  \end{eqnarray}
  If $\mathcal{T}=I$,
  \begin{eqnarray}
    &&
    \mathfrak{r}^T(E)=\mathfrak{r}(E),\quad \mathfrak{r}'^{T}(E)=\mathfrak{r}'(E), \quad
    \nonumber \\
    &&
    \mathfrak{t}^{T}(E)=\mathfrak{t}'(E).
  \end{eqnarray}
  Thus, the transmission $G_{R\rightarrow L}=\text{tr}[\mathfrak{t}'(E)\mathfrak{t}'^{\dagger}(E)]=\text{tr}[\mathfrak{t}^T(E)\mathfrak{t}^{*}(E)]=\text{tr}[\mathfrak{t}(E)\mathfrak{t}^{\dagger}(E)]=G_{L\rightarrow R}$.

\item PHS:
  \begin{eqnarray}
    \mathcal{P}K^T(E)\mathcal{P}^{-1}&=&W^{\dagger}\frac{1}{E+H}W=-K(-E),\\
    \mathcal{P}S^T(E)\mathcal{P}^{-1}&=&\frac{1+i\pi K(-E)}{1-i\pi K(-E)}=S^{-1}(-E).
  \end{eqnarray}
  If $\mathcal{P}=I$, we have $S^T(E)=S^{-1}(-E)$.

\item PHS$^{\dagger}$:
  \begin{eqnarray}
    \mathcal{P}K^*(E)\mathcal{P}^{-1}&=&W^{\dagger}\frac{1}{E^{*}+H}W=-K(-E^{*}),\\
    \mathcal{P}S^*(E)\mathcal{P}^{-1}&=&\frac{1-i\pi K(-E^*)}{1+i\pi K(-E^*)}=S(-E^*).
  \end{eqnarray}
  If $\mathcal{P}=I$, $S^*(E)=S(-E^*)$.

\item CS:
  \begin{eqnarray}
    \mathcal{C}K^{\dagger}(E)\mathcal{C}^{-1}&=&
                                                 W^{\dagger}\frac{1}{E^*+H}W=-K(-E^*),\\
    \mathcal{C}S^{\dagger}(E)\mathcal{C}^{-1}&=&\frac{1-i\pi K(-E^*)}{1+i\pi K(-E^*)}=S(-E^*).
  \end{eqnarray}
  If $\mathcal{C}=I$ and $E=0$,
  \begin{eqnarray}
    \mathfrak{t}'=\mathfrak{t}^\dagger,
  \end{eqnarray}
  so we have 
  \begin{eqnarray}
    G_{R\rightarrow L}=\text{tr}[\mathfrak{t}'\mathfrak{t}'^\dagger]=\text{tr}[\mathfrak{t}^\dagger \mathfrak{t}]=G_{L\rightarrow R}.
  \end{eqnarray}

\item SLS:
  \begin{eqnarray}
    \mathcal{L}K(E)\mathcal{L}^{-1} &=&
                                        W^\dagger \frac{1}{E+H}W=-K(-E),\\
    \mathcal{L}S(E)\mathcal{L}^{-1} &=&
                                        \frac{1+i\pi K(-E)}{1-i\pi K(-E)}=S^{-1}(-E).
  \end{eqnarray}
  If $\mathcal{L}=I$, $S(E)=S^{-1}(-E)$.

\item Pseudo-Hermiticity:
  \begin{eqnarray}
    \eta K^{\dagger}(E)\eta^{-1}&=&W^\dagger\frac{1}{E^*-H}W=K(E^*),\\
    \eta S^\dagger(E)\eta^{-1}&=&\frac{1+i\pi K(E^*)}{1-i\pi K(E^*)}=S^{-1}(E^*).
  \end{eqnarray}
  If $\eta=I$ and $\text{Im}(E)=0$, $S(E)$ is unitary, so $G_{R\rightarrow L}=G_{L\rightarrow R}$.

\end{enumerate}

\section{Winding number of exceptional points}
\label{app:winding}

\begin{figure}
\centering
\includegraphics[width=1\columnwidth]{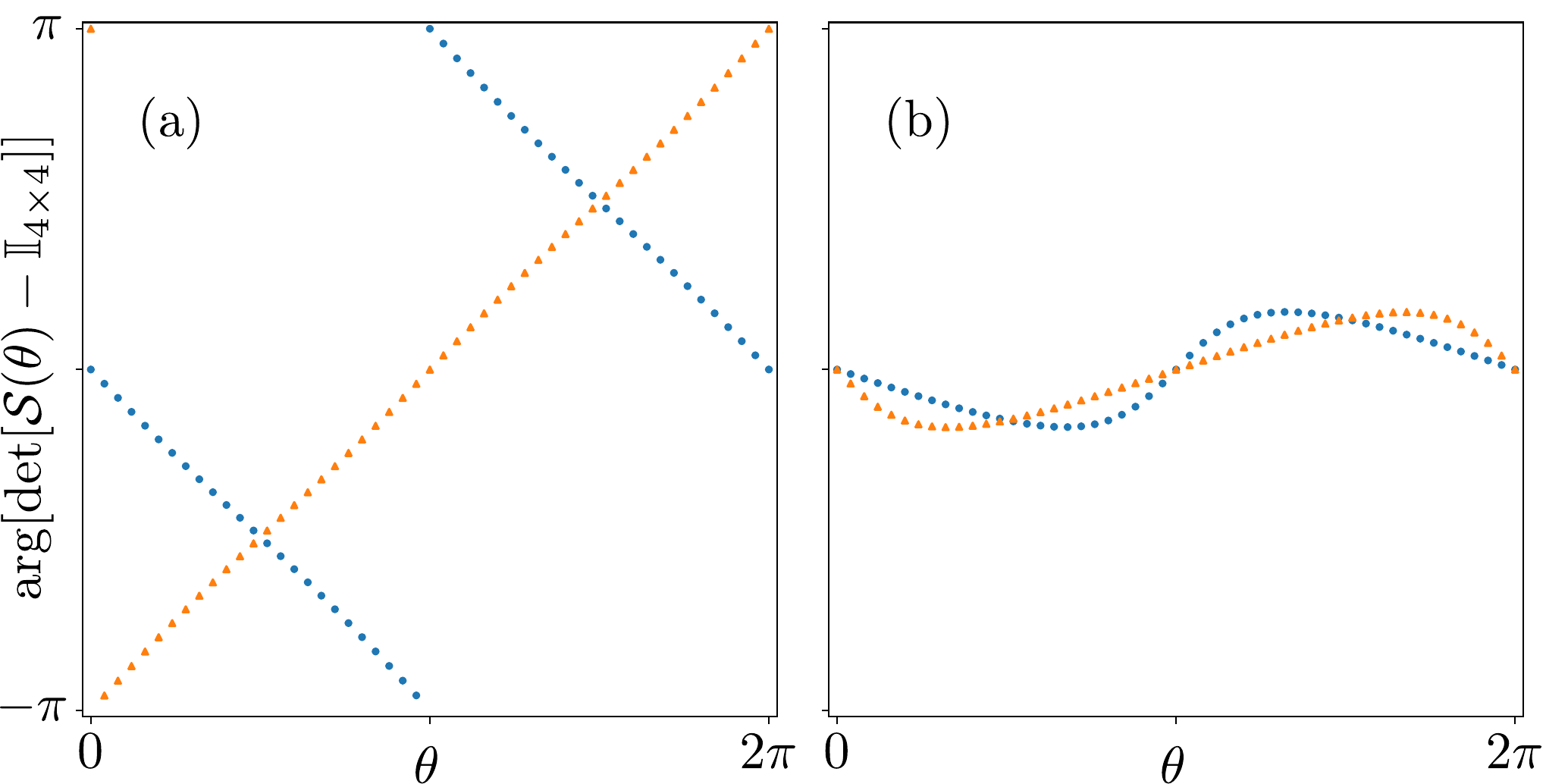}
\caption{$\text{arg}[\text{det}[\mathcal{S}(\theta)-\mathbb{I}_{4\times 4}]]$ as a function of $\theta$ (the parameter of the closed loop $s^1$), when encircling an EP (a) or not (b). 
Blue dots and orange triangles correspond to two different closed loops. 
The latter enclose one or the other EP in panel (a).
All the plots are with $\alpha=\pi/4$ and $\gamma=0.5$.
\label{fig: winding}}
\end{figure}

We calculate the winding number $\mathcal{W}$ in the polar coordinate with
\begin{eqnarray}
\mathcal{W}=\frac{1}{2\pi i}\int_{0}^{2\pi}d\theta\frac{d}{d\theta}\text{log}[\text{det}[\mathcal{S}(\theta)-\mathbb{I}_{4\times 4}]].
\end{eqnarray}
Here the reference point is $\varepsilon_{\textbf{k}_{\text{EP}}}=0$.
Choosing a closed circle $s^1$ encircling an EP leads to $\mathcal{W}=1$ for one EP and $-1$ for another EP~[see Fig.~\ref{fig: winding}(a)].
Thus to gap out one of EPs, the only way is to annihilate both of them simultaneously.

\section{Long-wavelength limit}
\label{app:longwavelength_limit}

We rewrite the non-unitary Ho-Chalker operator as
\begin{eqnarray}
\mathcal{S}(\mathbf{k})=\begin{pmatrix}
0&\tilde{M}(k_x)\\
\tilde{N}(k_y)&0
\end{pmatrix},
\end{eqnarray}
with 
\begin{eqnarray}
\tilde{M}(k_x)=\begin{pmatrix}
e^{-\gamma}&0\\
0&1
\end{pmatrix}\begin{pmatrix}
\sin\alpha e^{-ik_x} & \cos\alpha\\
-\cos\alpha & \sin\alpha e^{ik_x}
\end{pmatrix},
              \\
\tilde{N}(k_y)=\begin{pmatrix}
e^{\gamma}&0\\
0&1
\end{pmatrix}\begin{pmatrix}
\cos\alpha & \sin\alpha e^{ik_y}\\
-\sin\alpha e^{-ik_y} & \cos\alpha
\end{pmatrix}.
\end{eqnarray}
Then, to derive the long-wavelength model, we consider a two-step time evolution that gives $\mathcal{S}^2(\mathbf{k})=\text{diag}[\tilde{M}(k_x)\tilde{N}(k_y),\tilde{N}(k_y)\tilde{M}(k_x)]$~\cite{Ho1996}.
Without loss of generality, we focus on the block $\tilde{M}(k_x)\tilde{N}(k_y)$ with $\alpha=\pi/4+m$, $k_x=p_x$, and $k_y=p_y+\pi$.
Expanding it for small $p_x$, $p_y$, and $\alpha$ will give $\tilde{M}(k_x)\tilde{N}(k_y)\simeq 1-i\mathcal{H}$ with
\begin{eqnarray}
\mathcal{H}=p_x\frac{\sigma_{-}}{2}+(p_y-i\gamma)\frac{\sigma_{+}}{2}+m\sigma_y,
\end{eqnarray}
which is reminiscent of the low-energy continuum model of the non-Hermitian Chern insulator~\cite{Yao2018a,Shen2018a}. 

\section{Skin effect of the edge state}
\label{app:se_es}

\begin{figure}
\centering
\includegraphics[width=1\columnwidth]{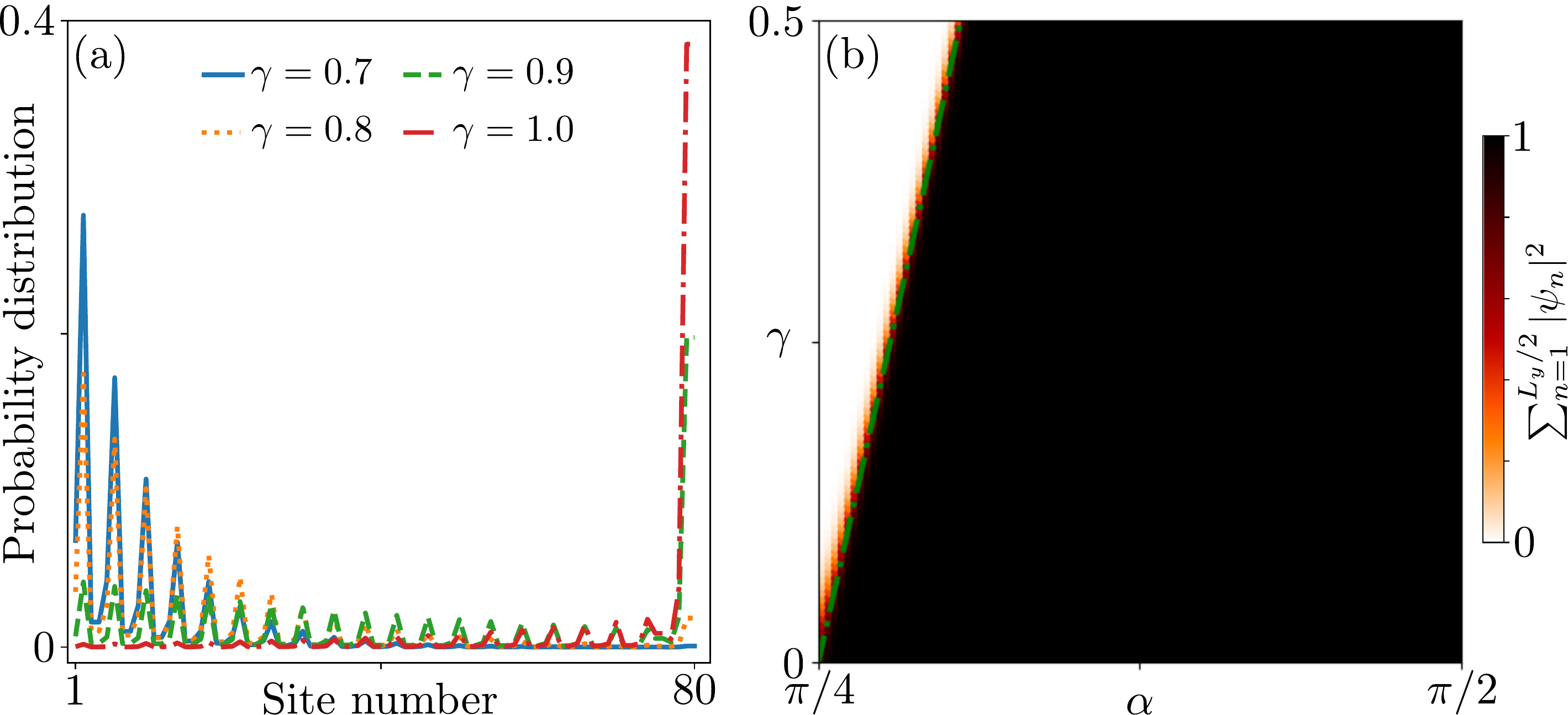}
\caption{(a) The probability distribution of the top edge state as a function of $\gamma$ with $\alpha=0.35\pi$ ($\gamma_{\text{EP}}\approx 0.937$) and a system size $L_y=20$ unit cells.
(b) The edge mode probability distribution is summed over half of the system, corresponding to unit cells with position $n\leq L_y/2$. This sum is plotted as a function of $\alpha$ and $\gamma$, using $L_y=80$. 
The blue dash-dotted line indicates the critical $\gamma$ for EPs, showing that the top edge mode is pushed to the bottom boundary as soon as EPs appear in the spectrum of the infinite system.
All plots are with $k_x=0.05$.
\label{fig: push_edge_state}}
\end{figure}

The edge state is not influenced by the non-Hermitian skin effect, due to its eigenphases being outside the region of winding number $\mathcal{W}=1$.
In this sense, a possible way to realize a skin effect for the boundary mode is to reshape such a region by increasing $\gamma$.
A criterion for this is the appearance of the EPs at $k_x=0$, where all the eigenphases $\varepsilon$ of the edge state are circled by those of the infinite system.
Therefore, once $\gamma>\gamma_{\text{EP}}$, the top boundary mode would be pushed to the bottom edge, see Fig.~\ref{fig: push_edge_state}.

\section{Higher-order skin effect}
\label{app:hose}

\begin{figure*}[tb]
\centering
\includegraphics[width=0.9\textwidth]{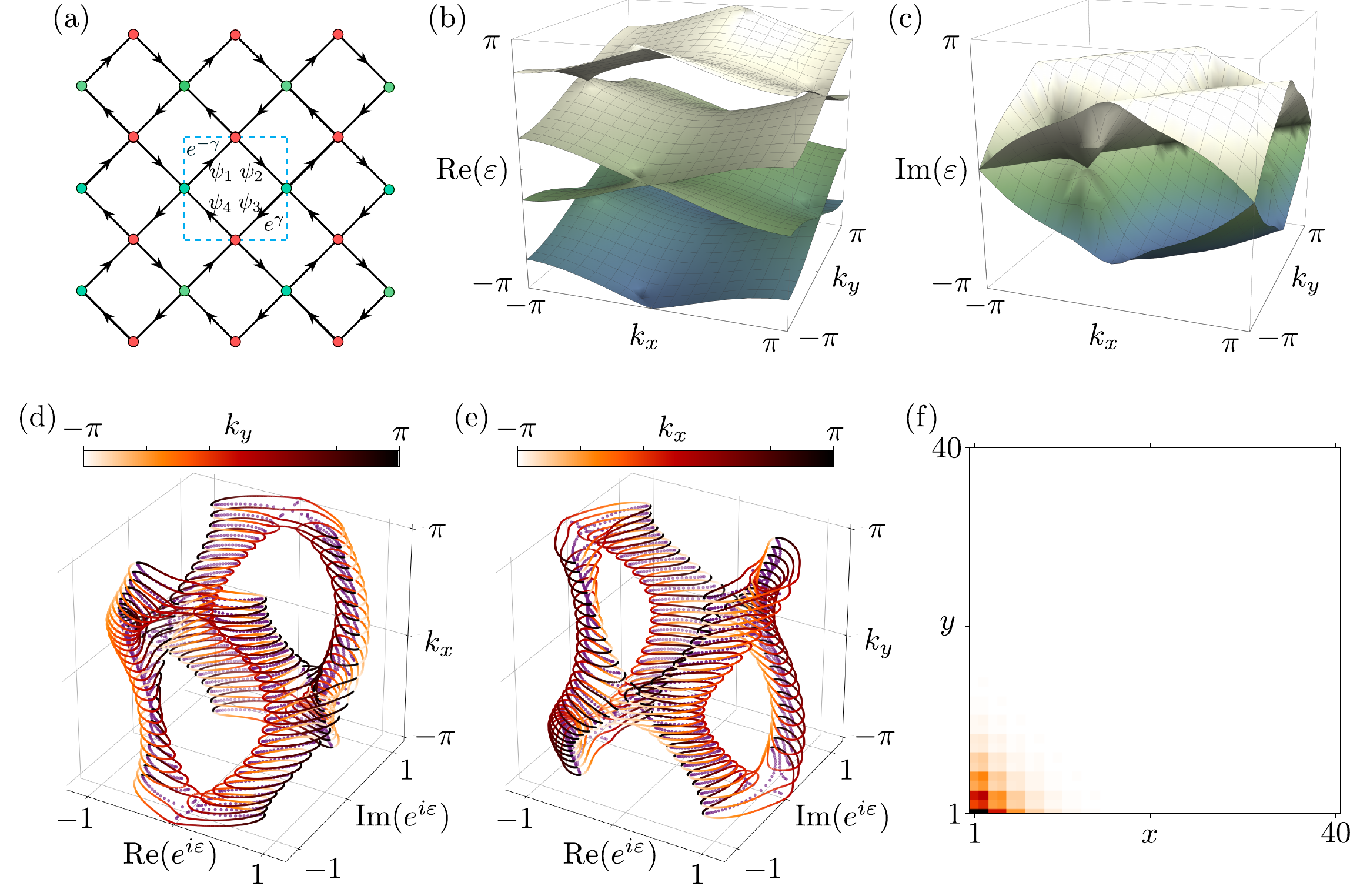}
\caption{(a) Illustration of the non-unitary network model with the second-order skin effect.
(b) and (c) are the real and imaginary eigenphase spectra. 
(d) and (e) are the ribbon geometry eigenphase spectrum encircled by the bulk dispersion along the $k_x$- and $k_y$-directions, respectively. 
(f) the probability distribution of all the eigenstates with a system size $20\times 20$. 
All of the plots are obtained using $\alpha=\pi/4$ and $\gamma=0.5$.
\label{fig: hose}}
\end{figure*}

By adding gain $e^{\gamma}$ to $\psi_3$ instead of $\psi_2$ [see Fig.~\ref{fig: hose}(a)], the non-unitary network also supports a second-order skin effect.
Because of $\text{det}[\mathcal{S}(\textbf{k})]=1$, the condition for EPs is still $e^{2i\varepsilon}=\pm 1$, and the EPs for $\varepsilon=0$ [shown in Fig.~\ref{fig: hose}(b-c)] are at
\begin{eqnarray}
  k_x=\pm\arccos(\frac{1}{\cos i\gamma \sin 2\alpha}),
  \quad k_y=\pi-k_x.
\end{eqnarray}  
In this sense, after the system becomes gapless at one high symmetry point $(0,\pi)$, these two EPs would move towards another high symmetry point $(\pi,0)$, but they can not be annihilated anymore.
An intuitive explanation is $1/|\cos i\gamma \sin 2\alpha|$ would be always larger than $0$.
Furthermore, with a ribbon geometry (along the $x$- or $y$-direction), the propagating modes can feel the amplification and suppression in opposite directions, which means a skin effect.
In Fig.~\ref{fig: hose}(d-e), the eigenphases of the ribbon are encircled by the eigenphase loops of the infinite system.
Next, if opening both the $x$- and $y$-direction, all modes are localized at the left bottom corner [see Fig.~\ref{fig: hose}(f)].
This is the so called second-order skin effect, which is an effect induced by gain and loss, instead of crystalline symmetries~\cite{Benalcazar61, Benalcazar_prb, Liu2021}.

\section{Scattering matrix for the Dirac point}
\label{app: sca_dirac}

For a slice of the network model which has $L_y=1$ but is infinite along the $x$-direction, such that $k_x$ is a good quantum number, the scattering matrix reads
\begin{eqnarray}\label{eq:Skx}
\hspace{-5mm}S_{\text{v, slice}}=\begin{pmatrix}
\frac{\cos\alpha-\sin\alpha e^{-ik_x}}{1-\cos\alpha\sin\alpha e^{-ik_x}} & -\frac{\sin\alpha\cos\alpha e^{-\gamma}}{1-\cos\alpha\sin\alpha e^{-ik_x}}\\
-\frac{\sin\alpha\cos\alpha e^{\gamma}}{1-\cos\alpha\sin\alpha e^{-ik_x}} &\frac{\sin\alpha e^{ik_x}-\cos\alpha}{1-\cos\alpha\sin\alpha e^{-ik_x}}
\end{pmatrix}.
\end{eqnarray}
Taking $\alpha\in (0,\pi/2)$, the Dirac point appears at momenta $(k_x,k_y)=(0,\pi)$, and values ($\alpha$, $\gamma$) that are obtained using Eq.~\eqref{eq:boundary_2}. For these values, Eq.~\eqref{eq:Skx} reveals a new conservation,
\begin{eqnarray}
\mathfrak{t}_{\text{v}}+\mathfrak{t}'_{\text{v}}=-2.
\end{eqnarray}
The latter can be proven by the transfer matrix of the slice, which reads
\begin{eqnarray}
\hspace{-5mm}M_{\text{v, slice}}=e^{\gamma}\begin{pmatrix}
e^{ik_x}-\frac{2}{\sin\alpha\cos\alpha}&\frac{1}{\sin\alpha}-\frac{e^{ik_x}}{\cos\alpha}\\
\frac{1}{\sin\alpha}-\frac{e^{-ik_x}}{\cos\alpha}&e^{-ik_x}-\frac{2}{\sin\alpha\cos\alpha}
\end{pmatrix}.\label{eq:m_unit}
\end{eqnarray}
Then the total transfer matrix can be obtained by the eigenvalue decomposition,
\begin{eqnarray}
M_{\text{v}}=U_{\text{evec}}\text{diag}[E_1^{L_y},E_2^{L_y}]U_{\text{evec}}^{-1},\label{eq:m_total}
\end{eqnarray}
where $U_{\text{evec}}$ is $2\times 2$ matrix composed of the eigenvectors of Eq.\ \eqref{eq:m_unit}. $E_1$ and $E_2$ are their corresponding eigenvalues,
\begin{eqnarray}
E_i=e^{\gamma}(\mu+(-1)^{i}\sqrt{\mu^2-1}),
\end{eqnarray}
with $\mu=\cos k_x-\frac{2}{\sin 2\alpha}$. 
It is obvious that there is also a conservation $E_1E_2=e^{2\gamma}$.
In Eq.\ \eqref{eq:m_unit}, if $(\phi_1, \phi_2)^T$ is an eigenvector for $E_i$ ($i=1,~2$), then $(\phi_2^{*}, \phi_1^{*})^T$ is also a solution for the same eigenvalue. 
Thus, it can be written as $\frac{1}{\sqrt{2}}(e^{i\varphi},1)^T$.
Since there is no degeneracy, another orthogonal eigenvector should be $\frac{1}{\sqrt{2}}(-e^{i\varphi}, 1)^T$.
Thus, Eq.\ \eqref{eq:m_total} now becomes
\begin{eqnarray}
M_{\text{v}}=\frac{1}{2}\begin{pmatrix}
E_1^{L_y}+E_2^{L_y}&e^{i\varphi}(E_1^{L_y}-E_2^{L_y})\\
e^{-i\varphi}(E_1^{L_y}-E_2^{L_y})&E_1^{L_y}+E_2^{L_y}
\end{pmatrix},
\end{eqnarray}
which leads to
\begin{eqnarray}
\mathfrak{t}_{\text{v}}=\frac{2(E_1E_2)^{L_y}}{E_1^{L_y}+E_2^{L_y}},
  \quad
\mathfrak{t}'_{\text{v}}=\frac{2}{E_1^{L_y}+E_2^{L_y}}.\label{eq:t_and_tp}
\end{eqnarray}
At the Dirac point $k_x=0$, $\mu=-\cosh\gamma$ leads to $E_1=-e^{2\gamma}$ and $E_2=-1$, which indicates
\begin{eqnarray}
\mathfrak{t}_{\text{v}}+\mathfrak{t}'_{\text{v}}=2\cdot (-1)^{L_y}.
\end{eqnarray}
And when $L_y$ goes to infinity,
\begin{eqnarray}
\lim_{L_y\rightarrow\infty}|\mathfrak{t}_{\text{v}}|&=&\lim_{L_y\rightarrow\infty}\frac{2e^{2\gamma L_y}}{e^{2\gamma L_y}+1}=2,\\
\lim_{L_y\rightarrow\infty}|\mathfrak{t}'_{\text{v}}|&=&\lim_{L_y\rightarrow\infty}\frac{2}{e^{2\gamma L_y}+1}=0.
\end{eqnarray}
When $k_x\neq 0$, $|\mu|>\cosh\gamma$ leads to
\begin{eqnarray}
|E_1|>e^{\gamma}(\sqrt{\cosh^2\gamma-1}-\mu)>e^{2\gamma}.
\end{eqnarray}
Correspondingly, $|E_2|<1$.
Then we have,
\begin{eqnarray}
\lim_{L_y\rightarrow\infty}|\mathfrak{t}_{\text{v}}|=0,~
\lim_{L_y\rightarrow\infty}|\mathfrak{t}'_{\text{v}}|=0.
\end{eqnarray}
Thus, the non-Hermitian Dirac point has a quantized transmission $4$ at $k_x=0$ and  $0$ transmission at $k_x\neq 0$.

\section{Path integral approach}
\label{Path integral approach}

In this Appendix, we provide an analysis of the network model using the functional integral representation of the (product of) resolvents (Green's functions).
The technique has been widely used in (disordered) Hermitian systems \cite{Efetov:1997fw}.

We start with the following non-Hermitian Dirac Hamiltonian 
\begin{align}
  \label{Dirac non-hermitian}
  \mathcal{H}(p)
  &
  = (p_x+i A_x) \sigma_z + (p_y + i A_y)\sigma_x
  \nonumber \\
  &\quad
  + m \sigma_y + i V \sigma_0.
\end{align}
Here, $A_{x,y}, m, V$ are arbitrary real function in space.
The Hamiltonian is the member of class D$^{\dag}$ and satisfies
\begin{align}
\mathcal{H}^*(p)= - \mathcal{H}(-p). 
\end{align}
When $A_x=V=0$, $A_y = \mathrm{const.}=-\gamma$, and $m = \mathrm{const}.$ the non-Hermitian Dirac Hamiltonian \eqref{Dirac non-hermitian} reduces to the continuum limit of the non-Hermitian Chern insulator model \eqref{dirac non hermitian Chern ins}.
In general, all these terms $A_{x,y}, V, m$ can have both uniform and inhomogeneous parts.
For example, if we consider non-Hermitian terms (gain or loss) for all four types links in the network model (Fig.\ \ref{fig: set_up}), $e^{\gamma_1}, e^{\gamma_2}, e^{\gamma_3}, e^{\gamma_4}$, the continuum limit of the Ho-Chalker time evolution operator is given by \eqref{Dirac non-hermitian} with $A_x \propto \gamma_1+\gamma_3$, $A_y \propto \gamma_2-\gamma_4$, and $V \propto \gamma_1+\gamma_2 -\gamma_3+\gamma_4$, while the mass term $m$ is still controlled by $\alpha$ entering in the vertex term.
This type of non-Hermitian Dirac Hamiltonians have been discussed, e.g., in 
\cite{2000PhRvL..84.1292L, 2002JPhA...35.2555B, Yao2018a, Shen2018a}.

We are interested in the Green's function, $G(z)= (z - \mathcal{H})^{-1}$ $(z\in \mathbb{C})$, and products thereof. 
With the class D$^{\dag}$ symmetry, the Green's function satisfies $G(\varepsilon+i\eta)^* = - G(-\varepsilon+ i\eta),$ where $\varepsilon$ and $\eta$ are the real and imaginary part of $z$, $z= \varepsilon+ i\eta$.
Hence, this symmetry relates the Green's functions at $z=\varepsilon+i \eta$ and $-\varepsilon+i\eta$.
In contrast, in Hermitian symmetry class D, the particle-hole symmetry relates the retarded and advanced Green's functions at $\varepsilon=0$.
This symmetry of the Green's functions can be used to reduce the number of functional integral variables;
it suffices to introduce a path integral only for the retarded sector, say, but not for the advanced one.
We should also note that in non-Hermitian systems the notion of retarded/advanced Green's function, distinguished by the small imaginary part of $z$, may not be sharply defined since the energy spectrum itself may be complex.
In the network model calculations of our interest, we have ideal leads attached to the non-Hermitian system, which may serve a role in selecting retarded/advanced Green's functions.

The Green functions can be represented by using functional integrals either over bosonic or fermionic fields defined on the two-dimensional space \cite{Efetov:1997fw}.
We shall consider the Gaussian functional integrals,
\begin{align}
  &
  Z_{F}^{\pm}
    =\int \mathcal{D}\left[\bar{\chi}_{\pm}, \chi_{\pm}\right]
    e^{ - S^{\pm}_F},
    \nonumber \\
  &
    S_F^{\pm}
    =
    -{i} \int d^2x\, \bar{\chi}_{\pm}
    \left(\varepsilon\pm {i} 0^+-\mathcal{H}\right) \chi_{\pm},
    \nonumber \\
  &
  Z_{B}^{\pm}
    =\int \mathcal{D}\left[\bar{\phi}_{\pm}, \phi_{\pm}\right]
    e^{ - S_B^{\pm}},
    \nonumber \\
  &
  S^{\pm}_B =
    -{i} \int d^2x\, \bar{\phi}_{\pm}
    \left(\varepsilon\pm {i}0^+-\mathcal{H}\right) \phi_{\pm},
\end{align}
where $\chi_{\pm}$ ($\phi_{\pm}$) are fermionic (bosonic) functional integral variables. 
The two-point correlation functions computed using either $S^{\pm}_{F}$ or $S^{\pm}_B$ reproduce the Green's functions $G(\varepsilon\pm i 0^+)$.
The products of Green's functions can be represented similarly by introducing flavors (multiple copies) of the functional integral variables.
Because of the identity $Z^{\pm}_F Z^{\pm}_{B} =1$, valid for any configurations of $A_{x,y}, m, V$, the fermionic and bosonic integrals can be combined when we perform quenched disorder averaging (the supersymmetry technique \cite{Efetov:1997fw}).
We will focus below on the product of one retarded and one advanced Green's function, and consider $S_F= S^+_F + S^-_F$ and $S_B= S^+_B + S^-_B$, with the total action $S=S_F+S_B$.

For convenience, we will work with the rotated basis: $(\sigma_z, \sigma_x, \sigma_y) \to (\sigma_y, \sigma_x, -\sigma_z)$.
In this basis, we can write 
\begin{align}
  \mathcal{H} &= (-2i) \sigma_y \left[ \left(
    \begin{array}{cc}
       \partial +  A &                      M \\
       \bar{M}       & \bar{\partial} +\bar{A}
    \end{array}
    \right) \right],
\end{align}
where we have introduced $(\partial_{x}\mp i \partial_y)/2 \equiv \partial (\bar{\partial}), (A_x \mp i A_y)/2 \equiv  - A (\bar{A}), (m\pm i V)/2 = M(\bar{M})$.
Introducing the left and right movers as
\begin{align}
  &
  \bar{\chi} (-\sigma_y\tau_z)
  = (\psi^{\dag}_{R},\psi^{\dag}_L)_{a=1,2},
   \quad
  \chi =
         (\psi^{\ }_{R},\psi^{\ }_L)^T_{a=1,2},
    \nonumber \\
    &
  \bar{\phi} (-\sigma_y\tau_z)
  = (\beta^{\dag}_{R},\beta^{\dag}_L)_{a=1,2},
   \quad
  \phi =
         (\beta^{\ }_{R},\beta^{\ }_L)^T_{a=1,2},
\end{align}
where $\tau_z$ acts on the retarded/advanced indices $\pm$, and rescaling $\psi^{\dag}\to \psi^{\dag}/\sqrt{2}$, $\psi\to \psi/\sqrt{2}$, etc., we obtain
\begin{align}
  \label{susy action}
  S_F  &=
     \int d^2x\, 
      \sum_{a=1}^2
    \Big[
     \psi^{\dag}_{aR}(\partial + A) \psi^{\ }_{aR}
    +
     \psi^{\dag}_{aL}(\bar{\partial}+\bar{A}) \psi^{\ }_{aL}
    \nonumber \\
  &\quad
    +
    M \psi^{\dag}_{aR} \psi^{\ }_{aL}
    +
    \bar{M}
     \psi^{\dag}_{aL} \psi^{\ }_{aR}
    \Big]
    \nonumber \\
  &\quad
    +
    \int d^2x\,
    \sum_{a} (\varepsilon - (-1)^a i 0^+)
    \left[
    \psi^{\dag}_R \psi^{\ }_L - \psi^{\dag}_L \psi^{\ }_R
    \right],
    \nonumber \\
    S_B  &=
     \int d^2x\, 
      \sum_{a=1}^2
    \Big[
     \beta^{\dag}_{aR}(\partial + A) \beta^{\ }_{aR}
    +
     \beta^{\dag}_{aL}(\bar{\partial}+\bar{A}) \beta^{\ }_{aL}
    \nonumber \\
  &\quad
    +
    M \beta^{\dag}_{aR} \beta^{\ }_{aL}
    +
    \bar{M}
     \psi^{\dag}_{aL} \beta^{\ }_{aR}
    \Big]
    \nonumber \\
  &\quad
    +
    \int d^2x\,
    \sum_{a} (\varepsilon - (-1)^a i0^+)
    \left[
    \beta^{\dag}_R \beta^{\ }_L - \beta^{\dag}_L \beta^{\ }_R
    \right].
\end{align}

When we set $0^+=0$, the total action $S=S_F+S_B$ enjoys ${\it GL}(2|2)$ symmetry.
Up to the term proportional to $0^+$, the action \eqref{susy action} is identical to the action used to study an Anderson localization problem
for Hermitian symmetry class BDI -- See e.g., Ref.\ \cite{2000NuPhB.583..475G} and (2.28) in Ref.\ \cite{2012PhRvB..85w5115R}.
In these papers, the Anderson localization problem of fermions hopping on a two-dimensional square lattice in the presence of background $\pi$-flux per plaquette and real random bipartite hopping elements was studied \cite{1997PhRvB..56.1061H}.
The relation between the non-Hermitian and Hermitian problems comes from the doubling or Hermitization \cite{1997NuPhB.504..579F, 1997PhRvB..56.1061H,1998PhRvL..80.4257M, 2000NuPhB.583..475G, Kawabata2019b, Luoxunlong2021}.
While the actions are identical, in the Hermitian class BDI problem, the flavor degrees of freedom (labeled by the index $a$) come from the valley degrees of freedom on the square lattice that arise due to the fermion doubling.
On the other hand, in the non-Hermitian class D$^{\dag}$ problem, the flavor degrees of freedom arise as we consider the product of two Green's functions.

Reintroducing now the term proportional to $0^+$, this term breaks the ${\it GL}(2|2)$ symmetry down to ${\it GL}(1|1)$. 
In the Hermitian class BDI problem, there is a similar term that again breaks ${\it GL}(2|2)$ symmetry, and also the $U(1)$ symmetry $\psi_{Aa}\to e^{i\theta}\psi_{Aa}$, $\psi^{\dag}_{Aa}\to e^{-i\theta}\psi^{\dag}_{Aa}$, $\beta_{Aa}\to e^{i\theta}\beta_{Aa}$, $\beta^{\dag}_{Aa}\to e^{-i\theta} \beta^{\dag}_{Aa}$ where $A=R/L$.
The scaling of this term under the renormalization group (RG) flow leads the Gade singularity in the density of states at the band center. 
On the other hand, since in the non-Hermitian class D$^{\dag}$ problem the terms proportional to $0^+$ and $\varepsilon$ are invariant under $U(1)$, it is not clear if we should expect the Gade singularity or alike.
(Note that in order to discuss the density of states, we don't have to introduce the retarded/advanced indices, and we get the field theory with ${\it GL}(1|1)$ symmetry.)

Assuming white noise distribution with zero mean for all disorder potentials, and setting $\varepsilon=0^+=0$, the weak coupling RG flow has been computed \cite{2000NuPhB.583..475G, 2012PhRvB..85w5115R}.
The gauge randomness is marginally relevant and decouples from the other coupling constants.
In particular, in the Hermitian class BDI problem, it does not affect the conductance.
When only $\mathrm{Re}\,M$ or $\mathrm{Im}\, M$ is non-zero, the system reduces to two copies of a Dirac fermion in Hermitian symmetry class D, perturbed by random mass perturbation.
The system flows to a clean critical point.
On the other hand, when both $\mathrm{Re}\, M$ and $\mathrm{Im}\, M$ are present, and their disorder strength are the same, a line of critical point is realized.
Along the critical line, for the case of Hermitian symmetry class BDI, the conductance changes continuously.

Finally, the functional integral approach formulated above can be used to derive the corresponding (super) quantum spin chain problem in one spatial dimension -- see, for example, Refs.\ \cite{PhysRevB.50.10788, 1996PMagL..73..145L, 1997NuPhB.497..639K} to see how this mapping works for the original Chalker-Coddington network model (the integer quantum Hall plateau transition).
In this mapping, one of the spatial directions is regarded as a fictitious time direction.
Being anisotropic, taking the vertical and horizontal directions as the time direction maps our non-Hermitian network model to different quantum spin chain problems. 
If the horizontal direction is regarded as a time direction, the resulting one-dimensional quantum system is non-Hermitian and nothing but the Hatano-Nelson model (in the presence of disorder).
This is consistent with the fact that the non-Hermitian skin effect is present in the $y$-direction in the non-Hermitian network problem.
On the other hand, if the vertical direction is regarded as a time direction, the resulting one-dimensional quantum system is Hermitian and the non-Hermiticity ($\equiv \gamma$) in the original network model language is mapped to a finite chemical potential, i.e., the system is tuned away from half-filling. 
This is consistent with the extended gapless phase with the EPs in the non-Hermitian network model language. 

\newpage

\bibliography{reference}
\end{document}